\documentclass[11pt,twocolumn,tighten]{aastex63}
\usepackage{hyperref}
\usepackage{color}
\usepackage{url}
\usepackage{natbib}
\usepackage{graphicx}
\usepackage{xfrac}
\usepackage{ulem}
\usepackage{mathtools}
\usepackage{CJK}
\usepackage{threeparttable}
\usepackage{multirow}
\usepackage{amsmath}

\begin{document}
\shorttitle{Multiphase CGM of DESI ELGs}
\shortauthors{Lan et al.}
\title{The Multiphase Circumgalactic Medium of DESI Emission-Line Galaxies at $z\sim1.5$} 
\correspondingauthor{Ting-Wen Lan}
\email{twlan@ntu.edu.tw}
\author[0000-0001-8857-7020]{Ting-Wen Lan}
\affiliation{Graduate Institute of Astrophysics, National Taiwan University, No. 1, Sec. 4, Roosevelt Rd., Taipei 10617, Taiwan}
\affiliation{Department of Physics and Center for Theoretical Physics, National Taiwan University, No. 1, Sec. 4, Roosevelt Rd., Taipei 10617, Taiwan}
\affiliation{Institute of Astronomy and Astrophysics, Academia Sinica, No. 1, Sec. 4, Roosevelt Rd., Taipei 10617, Taiwan}
\author[0000-0002-7738-6875]{J. Xavier Prochaska}
\affiliation{Department of Astronomy and Astrophysics, University of California at Santa Cruz, 1156 High Street, Santa Cruz, CA 95064, USA}
\affiliation{University of California Observatories, Lick Observatory, 1156 High Street, Santa Cruz, CA 95064, USA}
\affiliation{Kavli Institute for the Physics and Mathematics of the Universe, University of Tokyo, Kashiwa 277-8583, Japan}
\affiliation{Division of Science, National Astronomical Observatory of Japan,2-21-1 Osawa, Mitaka, Tokyo 181-8588, Japan}
\author{J.~Aguilar}
\affiliation{Lawrence Berkeley National Laboratory, 1 Cyclotron Road, Berkeley, CA 94720, USA}
\author[0000-0001-6098-7247]{S.~Ahlen}
\affiliation{Department of Physics, Boston University, 590 Commonwealth Avenue, Boston, MA 02215 USA}
\author[0000-0003-2923-1585]{A.~Anand}
\affiliation{Lawrence Berkeley National Laboratory, 1 Cyclotron Road, Berkeley, CA 94720, USA}
\author[0000-0001-9712-0006]{D.~Bianchi}
\affiliation{Dipartimento di Fisica ``Aldo Pontremoli'', Universit\`a degli Studi di Milano, Via Celoria 16, I-20133 Milano, Italy}
\affiliation{INAF-Osservatorio Astronomico di Brera, Via Brera 28, 20122 Milano, Italy}
\author{D.~Brooks}
\affiliation{Department of Physics \& Astronomy, University College London, Gower Street, London, WC1E 6BT, UK}
\author[0000-0001-7316-4573]{F.~J.~Castander}
\affiliation{Institut d'Estudis Espacials de Catalunya (IEEC), c/ Esteve Terradas 1, Edifici RDIT, Campus PMT-UPC, 08860 Castelldefels, Spain}
\affiliation{Institute of Space Sciences, ICE-CSIC, Campus UAB, Carrer de Can Magrans s/n, 08913 Bellaterra, Barcelona, Spain}
\author{T.~Claybaugh}
\affiliation{Lawrence Berkeley National Laboratory, 1 Cyclotron Road, Berkeley, CA 94720, USA}
\author[0000-0002-1769-1640]{A.~de la Macorra}
\affiliation{Instituto de F\'{\i}sica, Universidad Nacional Aut\'{o}noma de M\'{e}xico,  Circuito de la Investigaci\'{o}n Cient\'{\i}fica, Ciudad Universitaria, Cd. de M\'{e}xico  C.~P.~04510,  M\'{e}xico}
\author{P.~Doel}
\affiliation{Department of Physics \& Astronomy, University College London, Gower Street, London, WC1E 6BT, UK}
\author[0000-0003-4992-7854]{S.~Ferraro}
\affiliation{Lawrence Berkeley National Laboratory, 1 Cyclotron Road, Berkeley, CA 94720, USA}
\affiliation{University of California, Berkeley, 110 Sproul Hall \#5800 Berkeley, CA 94720, USA}
\author[0000-0002-3033-7312]{A.~Font-Ribera}
\affiliation{Institut de F\'{i}sica d’Altes Energies (IFAE), The Barcelona Institute of Science and Technology, Edifici Cn, Campus UAB, 08193, Bellaterra (Barcelona), Spain}
\author[0000-0002-2890-3725]{J.~E.~Forero-Romero}
\affiliation{Departamento de F\'isica, Universidad de los Andes, Cra. 1 No. 18A-10, Edificio Ip, CP 111711, Bogot\'a, Colombia}
\affiliation{Observatorio Astron\'omico, Universidad de los Andes, Cra. 1 No. 18A-10, Edificio H, CP 111711 Bogot\'a, Colombia}
\author[0000-0001-9632-0815]{E.~Gaztañaga}
\affiliation{Institut d'Estudis Espacials de Catalunya (IEEC), c/ Esteve Terradas 1, Edifici RDIT, Campus PMT-UPC, 08860 Castelldefels, Spain}
\affiliation{Institute of Cosmology and Gravitation, University of Portsmouth, Dennis Sciama Building, Portsmouth, PO1 3FX, UK}
\affiliation{Institute of Space Sciences, ICE-CSIC, Campus UAB, Carrer de Can Magrans s/n, 08913 Bellaterra, Barcelona, Spain}
\author{G.~Gutierrez}
\affiliation{Fermi National Accelerator Laboratory, PO Box 500, Batavia, IL 60510, USA}
\author[0000-0003-0201-5241]{R.~Joyce}
\affiliation{NSF NOIRLab, 950 N. Cherry Ave., Tucson, AZ 85719, USA}
\author[0000-0002-0000-2394]{S.~Juneau}
\affiliation{NSF NOIRLab, 950 N. Cherry Ave., Tucson, AZ 85719, USA}
\author{R.~Kehoe}
\affiliation{Department of Physics, Southern Methodist University, 3215 Daniel Avenue, Dallas, TX 75275, USA}
\author[0000-0003-3510-7134]{T.~Kisner}
\affiliation{Lawrence Berkeley National Laboratory, 1 Cyclotron Road, Berkeley, CA 94720, USA}
\author[0000-0001-6356-7424]{A.~Kremin}
\affiliation{Lawrence Berkeley National Laboratory, 1 Cyclotron Road, Berkeley, CA 94720, USA}
\author[0000-0003-1838-8528]{M.~Landriau}
\affiliation{Lawrence Berkeley National Laboratory, 1 Cyclotron Road, Berkeley, CA 94720, USA}
\author[0000-0001-7178-8868]{L.~Le~Guillou}
\affiliation{Sorbonne Universit\'{e}, CNRS/IN2P3, Laboratoire de Physique Nucl\'{e}aire et de Hautes Energies (LPNHE), FR-75005 Paris, France}
\author[0000-0003-4962-8934]{M.~Manera}
\affiliation{Departament de F\'{i}sica, Serra H\'{u}nter, Universitat Aut\`{o}noma de Barcelona, 08193 Bellaterra (Barcelona), Spain}
\affiliation{Institut de F\'{i}sica d’Altes Energies (IFAE), The Barcelona Institute of Science and Technology, Edifici Cn, Campus UAB, 08193, Bellaterra (Barcelona), Spain}
\author[0000-0002-1125-7384]{A.~Meisner}
\affiliation{NSF NOIRLab, 950 N. Cherry Ave., Tucson, AZ 85719, USA}
\author{R.~Miquel}
\affiliation{Instituci\'{o} Catalana de Recerca i Estudis Avan\c{c}ats, Passeig de Llu\'{\i}s Companys, 23, 08010 Barcelona, Spain}
\affiliation{Institut de F\'{i}sica d’Altes Energies (IFAE), The Barcelona Institute of Science and Technology, Edifici Cn, Campus UAB, 08193, Bellaterra (Barcelona), Spain}
\author[0000-0002-2733-4559]{J.~Moustakas}
\affiliation{Department of Physics and Astronomy, Siena University, 515 Loudon Road, Loudonville, NY 12211, USA}
\author[0000-0001-9070-3102]{S.~Nadathur}
\affiliation{Institute of Cosmology and Gravitation, University of Portsmouth, Dennis Sciama Building, Portsmouth, PO1 3FX, UK}
\author[0000-0002-0644-5727]{W.~J.~Percival}
\affiliation{Department of Physics and Astronomy, University of Waterloo, 200 University Ave W, Waterloo, ON N2L 3G1, Canada}
\affiliation{Perimeter Institute for Theoretical Physics, 31 Caroline St. North, Waterloo, ON N2L 2Y5, Canada}
\affiliation{Waterloo Centre for Astrophysics, University of Waterloo, 200 University Ave W, Waterloo, ON N2L 3G1, Canada}
\author[0000-0001-7145-8674]{F.~Prada}
\affiliation{Instituto de Astrof\'{i}sica de Andaluc\'{i}a (CSIC), Glorieta de la Astronom\'{i}a, s/n, E-18008 Granada, Spain}
\author[0000-0001-6979-0125]{I.~P\'erez-R\`afols}
\affiliation{Departament de F\'isica, EEBE, Universitat Polit\`ecnica de Catalunya, c/Eduard Maristany 10, 08930 Barcelona, Spain}
\author{G.~Rossi}
\affiliation{Department of Physics and Astronomy, Sejong University, 209 Neungdong-ro, Gwangjin-gu, Seoul 05006, Republic of Korea}
\author[0000-0002-9646-8198]{E.~Sanchez}
\affiliation{CIEMAT, Avenida Complutense 40, E-28040 Madrid, Spain}
\author{D.~Schlegel}
\affiliation{Lawrence Berkeley National Laboratory, 1 Cyclotron Road, Berkeley, CA 94720, USA}
\author{M.~Schubnell}
\affiliation{Department of Physics, University of Michigan, 450 Church Street, Ann Arbor, MI 48109, USA}
\affiliation{University of Michigan, 500 S. State Street, Ann Arbor, MI 48109, USA}
\author[0000-0002-6588-3508]{H.~Seo}
\affiliation{Department of Physics \& Astronomy, Ohio University, 139 University Terrace, Athens, OH 45701, USA}
\author[0000-0002-3461-0320]{J.~Silber}
\affiliation{Lawrence Berkeley National Laboratory, 1 Cyclotron Road, Berkeley, CA 94720, USA}
\author{D.~Sprayberry}
\affiliation{NSF NOIRLab, 950 N. Cherry Ave., Tucson, AZ 85719, USA}
\author[0000-0003-1704-0781]{G.~Tarl\'{e}}
\affiliation{University of Michigan, 500 S. State Street, Ann Arbor, MI 48109, USA}
\author{B.~A.~Weaver}
\affiliation{NSF NOIRLab, 950 N. Cherry Ave., Tucson, AZ 85719, USA}
\author[0000-0001-5381-4372]{R.~Zhou}
\affiliation{Lawrence Berkeley National Laboratory, 1 Cyclotron Road, Berkeley, CA 94720, USA}
\author[0000-0002-6684-3997]{H.~Zou}
\affiliation{National Astronomical Observatories, Chinese Academy of Sciences, A20 Datun Road, Chaoyang District, Beijing, 100101, P.~R.~China}

\begin{abstract}
We study the multiphase circumgalactic medium (CGM) of emission line galaxies (ELGs) at $z\sim1.5$, traced by MgII$\lambda2796$, $\lambda2803$ and CIV$\lambda1548$, $\lambda1550$ absorption lines, using approximately 8,000
ELG-quasar pairs from the Dark Energy Spectroscopic Instrument. Our results show that both the mean rest equivalent width ($W_{0}$) profiles and covering fractions of MgII and CIV increase with ELG stellar mass at similar impact parameters, but show similar distributions when normalized by the virial radius. Moreover, warm CIV gas has a more extended distribution than cool MgII gas. 
The dispersion of MgII and CIV gas velocity offsets relative to the galaxy redshifts rises from $\sim100 \, \rm km \, s^{-1}$ within halos to $\sim 200 \, \rm km \, s^{-1}$ beyond.
We explore the relationships between MgII and CIV $W_{0}$ and show that the two are not tightly coupled: at a fixed absorption strength of one species, the other varies by several-fold, indicating distinct kinematics between the gas phases traced by each. 
We measure the line ratios, FeII/MgII and CIV/MgII, of strong MgII absorbers and find that at $<0.2$ virial radius, the FeII/MgII ratio is elevated, while the CIV/MgII ratio is suppressed compared with the measurements on larger scales, both with $\sim4\, \sigma$ significance. 
We argue that multiphase gas that is not co-spatial is required to explain the observational results. Finally, 
by combining with the measurements in the literature, we show the redshift evolution of CGM properties and estimate the neutral hydrogen, metal, and dust masses in the CGM of DESI ELGs --- found to be comparable to those in the ISM.

\end{abstract}

\section{Introduction}
The circumgalactic medium (CGM) is a multiphase reservoir of gas surrounding galaxies, spanning a wide range of densities and temperatures, and carrying vital information about galaxy evolution \citep[see][for a review]{Tumlinson17}. Gas inflows and outflows regulate galaxy growth and leave observational signatures in the CGM \citep[See][for a review]{Faucher23}. Via absorption line spectroscopy, the properties of the CGM have been probed with various absorption line species. For example, MgII is used to trace cool gas ($\sim10^{4}$ K) \citep[e.g.,][]{Bergeron86, Kacprzak10, chen10, zhu13, lan14, lan17, anand21, napolitano23, chang24, shaban25, Chang2026},  while CIV ($\sim10^{4.5}$–$10^{5}$ K) and OVI ($\sim10^{5.5}$ K) are used to probe gas in warmer phases \citep[e.g.,][]{chen01, tumlinsonovi, Bordoloi14, Liang14, Kacprzak15, burchett16, Anand25, garza25}. Over the past decade, observations with the COS instrument on Hubble Space Telescope have characterized the multiphase nature of the CGM at low redshifts $z<0.4$ \citep[e.g.,][]{tumlinson13, werk2013, stocke13, borthakur16, heckman17, prochaska17, burchett19, ng19, zahedy21, tchernyshyov22, klimenko23, smailagic23, mishra24}. Extending these studies to higher redshifts, however, remains challenging, as it requires both precise galaxy redshift measurements and access to multiple absorption tracers.

At $z>1.4$, MgII and CIV absorption lines tracing cool and warm gas shift into the optical wavelength regions, making ground-based studies feasible \citep[e.g.,][]{Steidel10, QPQ14, turner14, Dutta21, Schroetter21, galbiati23}. Nevertheless, constructing large galaxy-quasar samples at these redshifts demands extensive observational resources for obtaining the galaxy redshift information and therefore the multiphase CGM studies at this cosmic epoch are still limited. 
One can now overcome this challenge with large spectroscopic datasets provided by ongoing and upcoming cosmological surveys, such as the Dark Energy Spectroscopic Instrument \citep[DESI,][]{levi13, desi22}, Prime Focus Spectrograph \citep[PFS,][]{pfs14} and Euclid \citep[][]{euclid22}. These surveys provide redshift measurements of millions of star-forming galaxies, enabling the assembly of large galaxy-quasar pair samples for multiphase CGM studies at $z>1.4$.

In this work, we utilize the latest DESI dataset to investigate the multiphase CGM of star-forming galaxies at $z\sim1.5$. We construct a sample of several thousand galaxy-quasar pairs within 300 kpc, measure MgII and CIV absorption strengths and gas kinematics, and explore correlations between CGM and galaxy properties. Our results offer new insights into the physical state of the multiphase CGM at high redshift and provide constraints on galaxy formation and feedback processes.

This paper is organized as follows: Section 2 describes the data and analysis, Section 3 presents results, Section 4 discusses implications, and Section 5 summarizes. We adopt a flat $\Lambda$CDM cosmology with $h=0.6766$ and $\Omega_m=0.30966$ \citep{planck18}. Magnitudes are in AB systems with Galactic extinction corrected \citep{SFD98}.

\section{Data analysis}
\subsection{Data}
To probe the CGM of galaxies via absorption lines, we use the spectroscopic dataset from the Dark Energy Spectroscopic Instrument (DESI) \citep{levi13,desi16,desi22}. In short, DESI is a Stage-IV cosmological survey, aiming to map the large-scale structure of the Universe across cosmic time and to obtain precise measurements for constraining the cosmological parameters. Utilizing the 4-m Mayall Telescope on the Kitt Peak Mountain with a spectroscopic instrument having 5000 fibers \citep{Silber23, miller24, poppett24}, DESI is designed to primarily observe four types of sources, (1) bright galaxies \citep{bgs23} and Milky Way stars \citep{mw23} during the bright time, (2) luminous red galaxies \citep[LRGs,][]{lrg23} at $0.4<z<1.2$, selected based on their colors and brightness, (3) star-forming galaxies that produce strong emission lines, emission-line galaxies \citep[ELGs,][]{elg23}, at $0.6<z<1.63$, selected based on their colors and brightness, and (4) quasars, selected based on a random forest algorithm, across a wide range of redshifts \citep{qso23}. Dedicated algorithms are developed to select targets \citep{target23}, assign fibers \citep{fiberinpre}, and incorporate them into the survey plan \citep{survey23}. After the observations, a dedicated and automatic pipeline is used to reduce and calibrate the raw data \citep{pipeline23}. The redshifts and the spectral types of the sources are then determined by the Redrock algorithm \citep{redrockinpre,Anand_redrock}. 
The spectral resolution of DESI ranges from 2500 (short wavelength) to 5000 (long wavelength) with a 0.8$\rm \, \AA$ pixel sampling \citep{pipeline23}.

In this work, we use the DR2 DESI datasets of ELGs and quasars which include data from the first 3-year DESI observations and are constructed following the method used in DR1 dataset \citep{lss23,dr1_datarelease}\footnote{Conveners: John Moustakas and Carlos Allende-Prieto}.
The cosmological measurements based on DR2 and DR1 are reported in \citet{dr2bao}\footnote{Conveners: Enrique Paillas, Sesh Nadathur, Mustapha Ishak and Willem Elbers} (baryon acoustic oscillations and cosmological constraints), \citet{dr2baof}\footnote{Conveners: Paul Martini, Andrei Cuceu, Eric Armengaud} (baryon acoustic oscillations from the Lyman alpha forest) and  \citet{dr1_cosmology}\footnote{Conveners: Eva-Maria Mueller, Dragan Huterer and Mustapha Ishak} (full-shape analysis).
More specifically, we use the following selections: 
\begin{itemize}
    \item \textbf{ELGs}: we select ELGs at $1.38< z_{\rm ELG}<1.62$ in order to detect MgII and CIV absorption lines in the DESI optical wavelength coverage. The lower and upper bounds of the redshifts are restricted by the rest-frame wavelength of CIV absorption and [OII] emission lines of galaxies respectively. In addition, we select ELGs satisfying $\rm log_{10}S/N(F_{OII})>0.9-0.2\times log_{10}\Delta \chi^{2}$, which combines [OII] emission line signal-to-noise ratio ($\rm S/N(F_{OII})$) and the $\Delta \chi^{2}$ value from Redrock \citep{elg23}. This selection is expected to yield a ELG sample with redshift purity $>99\%$ \citep{elg23, vi23}. With these two selections, the ELG sample consists of $1,600,958$ galaxies. 
    \item \textbf{Quasars}: we select sources in the LSS catalog that are identified as quasars by both the Redrock ($SPECTYPE=QSO$) and the QuasarNet pipelines \citep{qsonet25} and only use sources with $\Delta \chi^{2}>30$ from the Redrock pipeline. This yields a quasar sample, consisting of $2,088,305$ quasars, with a high purity \citep{qso23, viqso23}. With the selection, Redrock and QuasarNet provide consistent redshifts. We use the quasar redshifts provided by the Redrock pipeline.
\end{itemize}
Using the ELG and quasar samples, we construct a sample of ELG-quasar pairs with $c (z_{\rm QSO}-z_{\rm ELG})/(1+z_{\rm ELG})>5000 \, \rm km \, s^{-1}$ to avoid 
so-called associated systems \citep[e.g.,][]{Shen12} and with physical impact parameters, $r_{p}$, complete to 300 kpc and 2 $r_{vir}$
at the redshifts of ELGs. 
We apply a selection on the signal-to-noise ratios (S/N) of the quasar spectra and only retain spectra with median S/N higher than 2 around MgII 
and CIV spectral regions at the rest-frame of galaxies. For the CIV measurements, an additional redshift cut of background quasar redshifts, $\rm 1216\times(1+z_{QSO})<1540\times(1+z_{ELG})$, is applied to avoid contamination by the 
Lyman-$\alpha$ forest. 
This ELG-quasar pair sample consists of 8,695 pairs with a median redshift of 1.48.

\subsection{Absorption line detection algorithm}
In order to detect and measure the absorption line properties from the background quasar spectra, we first use the non-negative matrix factorization (NMF) technique implemented by \citet[][]{zhunmf16} to model and normalize the intrinsic quasar spectral features. This technique has been adopted in previous absorption line studies \citep[e.g.,][]{zhu13,chang24,ng25}. We further use a median filter with 85 pixels to account for the remaining small-scale fluctuation that was not captured by the NMF reconstruction \citep{napolitano23, chang24}. We obtain the final normalized spectra by normalizing the original DESI quasar spectra by their NMF reconstruction and the median-filter smoothed spectra. 

With the final normalized spectra, we search for absorption line features with a matched filter technique. We use the composite spectra of MgII absorbers from \citet{lan17}, select the MgII and CIV absorption line regions as the matched filter templates, and scan through the spectral window within $\rm -600 \, km \, s^{-1}$ and $\rm 600 \, km \, s^{-1}$ of the redshifts of ELGs. We estimate the corresponding uncertainty by running the matched filter through the error arrays of individual spectra. We retain spectra with pixels having matched-filter $\rm S/N>3$ across the spectral range for follow-up absorption line search.

For the spectra with matched-filter $\rm S/N>3$, we use double Gaussian profiles with a fixed separation of the MgII and CIV doublets to fit the absorption features using the velocity at the maximum S/N as the initial velocity. We obtain the best-fit rest equivalent widths, $W_{0}$, of the two lines of MgII$\lambda2796, \lambda2803$ and CIV$\lambda1548, \lambda1550$, their intrinsic line width\footnote{The DESI spectral resolution is included in the fitting.} $\sigma$, the relative central velocities of the absorption lines with respect to the redshifts of ELGs, $v_{c}$, and the uncertainties of all parameters. 
Finally, we consider (1) MgII absorption lines with the S/N ($W_{0}^{\lambda2796}/\sigma_{W_{0}^{\lambda2796}}$) of $W_{0}^{\lambda2796}$  $>4$ and S/N ($W_{0}^{\lambda2803}/\sigma_{W_{0}^{\lambda2803}}$)  of $W_{0}^{\lambda2803}$ $>2$ and the line ratio $0.2<\frac{W_{0}^{\lambda2803}}{W_{0}^{\lambda2796}}<2$ are detected sources and similarly (2) CIV absorption lines with S/N of $W_{0}^{\lambda 1548}>4$ and $W_{0}^{\lambda 1550}>2$ are detected sources. 

We find that some MgII absorption lines show distinct two absorption line components. A single set of double Gaussian profiles is not sufficient to capture the entire absorption profiles. To identify and measure the total $W_{0}$ of those systems, we also perform the fitting using two sets of double Gaussian profiles to all MgII detected sources based the fitting results with a single set of double Gaussian profiles. For MgII systems with the two components both having (1) S/N of $W_{0}^{\lambda2796}>3$ and $W_{0}^{\lambda2803}>1.5$ and (2) the sum of the $W_{0}^{\lambda2796}$ of the two components being larger than the origin $W_{0}^{\lambda2796}$ by $\sigma_{W_{0}^{\lambda2796}}$, we adopt the sum of the two $W_{0}^{\lambda2796}$ as the $W_{0}^{\lambda2796}$ of the systems and the $W_{0}$-weighted velocities of the two components as the central velocities. To estimate the line width $\sigma$ of the those systems, we use a Gaussian kernel with 68 $\rm km\, s^{-1}$ (standard deviation) to smooth the spectra, fit the smoothed spectra with a single set of double Gaussian profiles, and estimate the line width by subtracting the kernel size from the measured line width $\sigma$ in quadrature. The properties of approximately $6\%$ of detected MgII absorbers are updated in this process. 

Finally, we only consider absorption line systems with the absolute velocity offsets with respect to the galaxies smaller than 500 $\rm km \, s^{-1}$. Pixels which are masked by the pipeline are excluded when performing the fitting. We also exclude quasar spectra with more than 5 pixels being masked by the pipeline 
within $5 \, \rm \AA$ of the two lines of MgII and CIV. We note that MgII and CIV are detected independently.

\begin{table}[]
\center
\caption{Number of detected absorption systems and total number of sightlines, complete to 300 kpc and 2$r_{vir}$}
\begin{tabular}{c|cc|cc}
\hline
                                         & \multicolumn{2}{c|}{MgII}  & \multicolumn{2}{c}{CIV}  \\
                                         \hline
                       & Detection & Total & Detection & Total  \\         
                         \hline
Spectral $S/N>2$    & 730 & 8620 &  524 & 6657    \\
\hline
Spectral $S/N>5$ & 665 & 6827 & 359 & 2114 \\
\hline
\end{tabular}
\label{table:sources}
\end{table}

To account for non-detection due to low S/N of the spectra, following the approach adopted in \citet{chang24}, we construct 100,000 mock spectra with a range of S/N and add absorption lines into the spectra with the line properties randomly selected from the detected absorption line systems. We then identify detected absorption systems from these mock spectra by running the matched filter scanning and line fitting analysis.
With this information, we estimate the detection rate of absorption lines with a given $W_{0}$ as a function of spectral S/N. For each detected absorber from the observed spectrum, given its absorption strength and the S/N of the spectrum, we obtain its weight $w$ by taking the inverse of the detection rate. The weight values are used when calculating the MgII and CIV covering fraction around ELGs. We show the estimated detection rates as a function of $W_{0}$ and S/N in Appendix~\ref{sec:com}.

For the spectra with matched-filter $\rm S/N<3$, we still perform the Gaussian fitting with the initial velocity set at 0 $\rm km \, s^{-1}$. This enables capturing possible absorption signals with lower S/N. These measurements are included when calculating the mean $W_{0}$ as a function of $r_{p}$ in Section~\ref{text:distribution}.

Table~\ref{table:sources} summaries the number of detected absorption systems  and the total number of quasar sightlines. In this work, we mainly use the sightlines with spectral $S/N>5$ unless otherwise mentioned.

\begin{figure*}
\center
\includegraphics[width=0.9\textwidth]{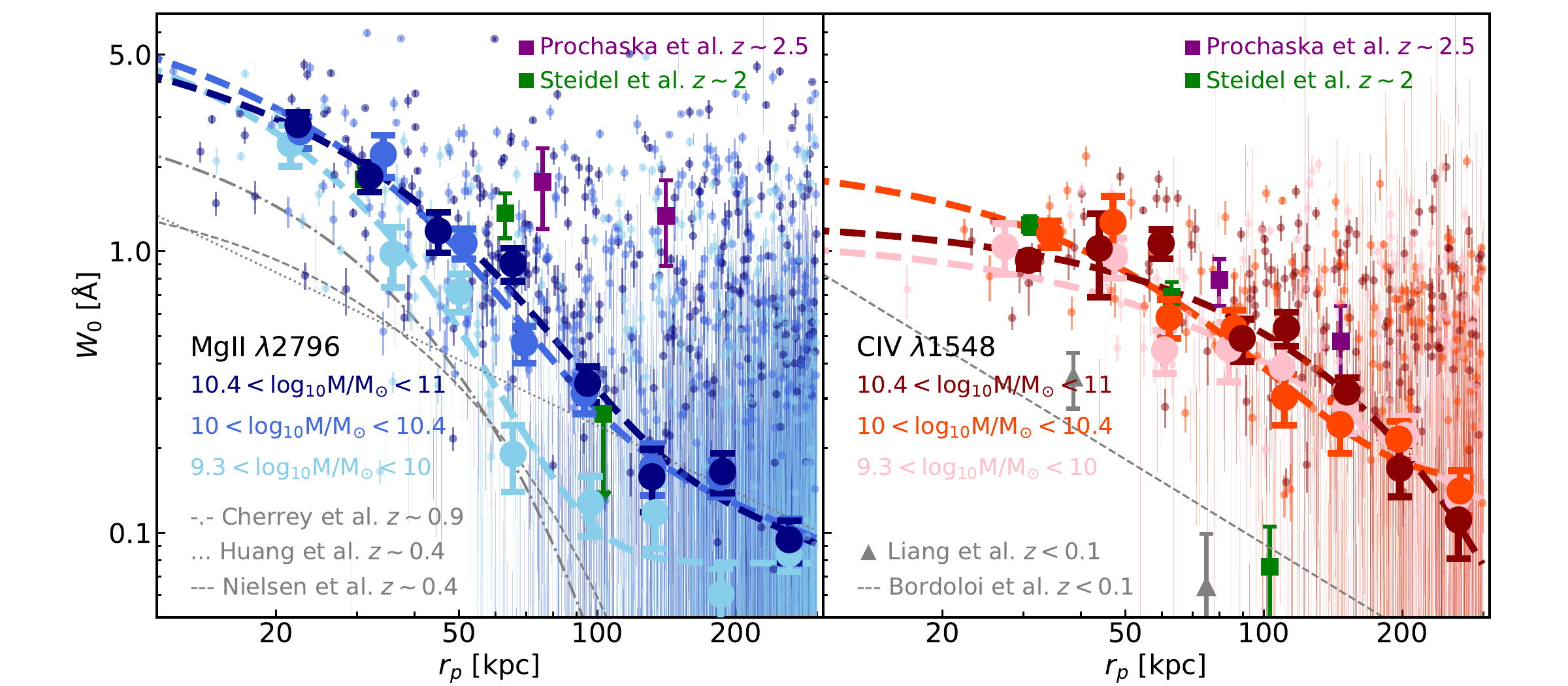}
\includegraphics[width=0.9\textwidth]{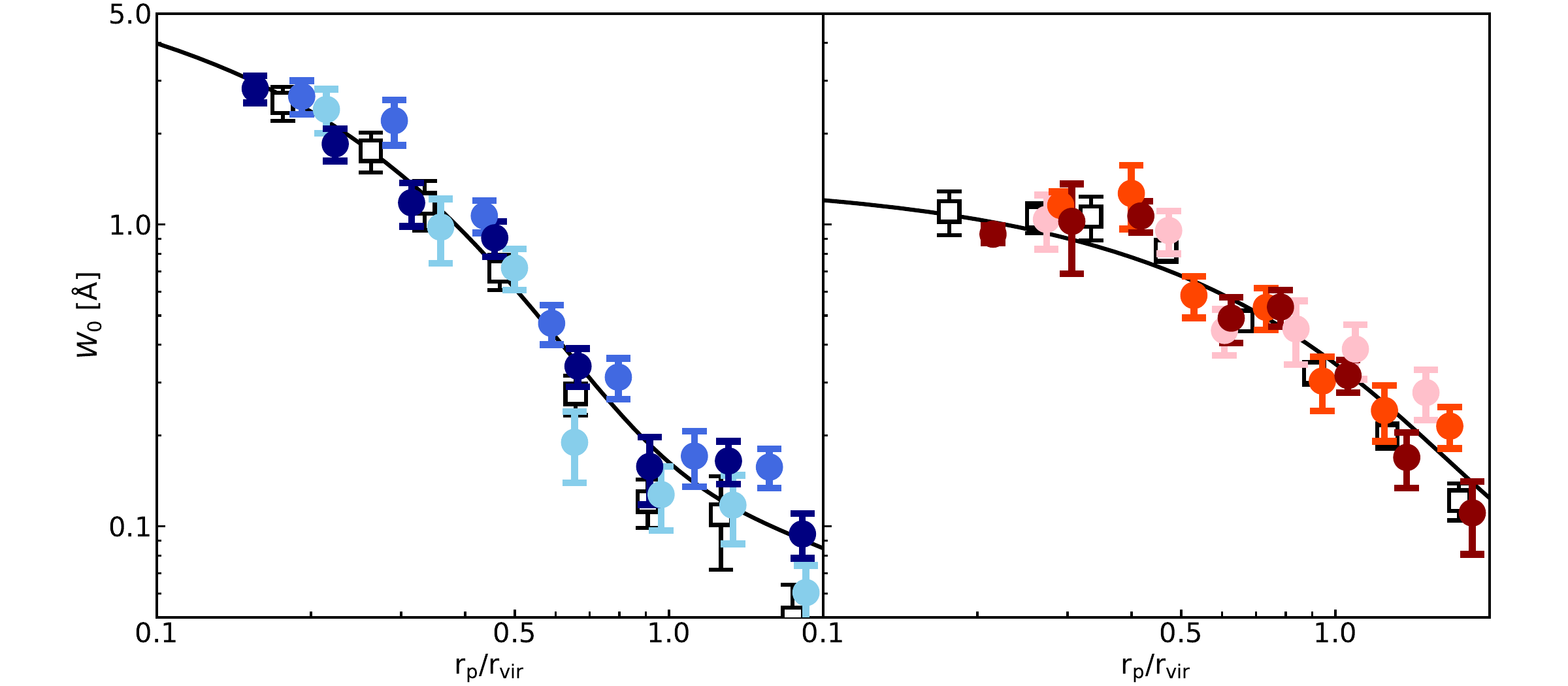}
\caption{Rest equivalent width distributions around ELGs. The upper panels show the rest equivalent width distribution in physical space and the lower panels show the rest equivalent width distribution in halo space $r_{p}/r_{vir}$. The left and right panels show the results for MgII and CIV respectively. The colors indicate measurements around DESI ELGs with different stellar mass as listed in the figure. The large data points show the mean rest equivalent widths from individual measurements with uncertainty based on bootstrapping the sample 500 times. The blue dashed lines and the red dashed lines show the best-fit functions (Equation 1) for MgII and CIV respectively. 
The black solid lines are the best-fit functions in halo space $r_{p}/r_{vir}$ (Equation 2).
In the upper left panel, the grey dashed, dotted, and dashed-dotted lines show the best-fit functions of MgII profiles from \citet{Nielsen13} and \citet{Huang21} at $z\sim0.3-0.4$  and \citet{Cherrey2025} at $z\sim0.9$ respectively. In the upper right panel, the grey dashed and data points show the CIV measurements at $z<0.1$ from \citet{Bordoloi14} and \citet{Liang14} respectively. The purple and green data points in the two panels are measurements at $z\sim2-2.5$ from \citet{QPQ14} and \citet{Steidel10}. CII absorption strength is used to estimate the expected MgII absorption strength. 
In the lower panels, the open data points are mean rest equivalent widths of MgII and CIV measured from mean composite spectra as a function of $\rm r_{p}/r_{vir}$.
}
\label{fig:MgII_CIV_REW_mass_dependence}
\end{figure*}

\subsection{Physical properties of ELGs}
For each ELG, we estimate three physical properties: stellar mass, star-formation rate (SFR), and the virial radius of the halo:
\begin{itemize}
    \item \textbf{Stellar mass $M_{*}$}: We use the CIGALE spectral energy distribution (SED) fitting package \citep{cigale} to estimate the stellar mass of ELGs. We include g, r, z, and WISE 1 observed magnitudes and their uncertainty from 
    the DESI Legacy Imaging Surveys \citep{dey19} and WISE \citep{wright10}
    together with the ELG redshifts. We adopt \citet{bc03} stellar population synthesis, Chabrier initial mass function \citep{chabrier03} and delayed star-formation history (with optional exponential burst). The typical statistical uncertainty of $M_{*}$ from CIGALE is $\sim0.15-0.2$ dex. 
    In this work, we focus on ELGs with stellar mass from $10^{9.3}$ to $10^{11} \, M_{\odot}$.
    \item \textbf{SFR}: We also use the SFR provided by CIGALE. The median SFR of the entire sample is $\sim 22 \,\rm  M_{\odot}\, yr^{-1}$, which is similar to the estimated SFR of DESI ELGs at similar redshifts based on multi-wavelength measurements from the COSMOS survey \citep{weaver22, lan24}. The typical uncertainty of SFR from CIGALE is $13 \, M_{\odot}\, yr^{-1}$.
    \item \textbf{Virial radius $r_{vir}$}: We estimate the halo mass for each ELG based on the stellar mass and halo mass relation from \citet{Behroozi13} and use the formula from \citet{Bryan98} to estimate the virial radius of the halo. We note that the statistical uncertainties in $M_{*}$ translate to approximately $10\%$ uncertainties in the virial radius.
\end{itemize}

\section{Results}
With this large sample of ELG and quasar pairs at $z\sim1.5$, we now explore the properties of the gas traced by MgII and CIV as a function of galaxy properties. The investigation is separated into two parts:
\begin{enumerate}
    \item In the first part, we explore the gas radial distribution around galaxies (Section~\ref{text:distribution}) and their kinematics (Section~\ref{text:kinematics}). The behaviors of the gas traced by MgII and CIV are investigated separately and independently. 
    \item In the second part, we explore the relationships between absorption strengths and relative kinematics of MgII and CIV absorption lines (Section~\ref{text:MgIIvsCIV}) and investigate how the relationships depend on the distances from galaxies (Section~\ref{text:lineratio}).
\end{enumerate}

\subsection{Gas distributions as a function of galaxy properties}
\label{text:distribution}
In this section, we explore the MgII and CIV gas radial distribution around galaxies as a function of galaxy properties.
\subsubsection{Stellar mass and halo mass dependence}
\textbf{Rest equivalent width distribution:}
We first explore the rest equivalent width distribution of MgII and CIV as a function of impact parameters. The upper left and right panels of Figure~\ref{fig:MgII_CIV_REW_mass_dependence} show the results of MgII $W_{0}^{\lambda2796}$ and CIV $W_{0}^{\lambda1548}$ respectively. The individual data points show the detected systems and the faint vertical bars are non-detected sightlines with best-fit measurements from the spectra and their uncertainty. The large data points show the mean $W_{0}$ as a function of $r_{p}$, which include both detected and non-detected sources. We note that for the non-detected sources, they still have measured $W_{0}$ values which can be positive or negative and mostly reflect the noise level of the spectra. The colors indicate the measurements for ELGs with different stellar mass as labeled in the figure.

\begin{table*}[]
\center
\caption{
Best-fit parameters for $W_{0}$ as a function of stellar mass
}
\begin{tabular}{ll|llll}
                      &                                     & $A_{w}$             &  $r_{w}$   &    $B_{w}$     & $\beta_{w}$          \\
                      \hline
\multirow{3}{*}{MgII} & $10^{9.3}<M_{*}<10^{10} M_{\odot}$  & 

$8.18\pm2.63$ & $17.26\pm2.20$ & $0.08\pm0.01$ & $0.00\pm0.00$  \\
                     
                      & $10^{10}<M_{*}<10^{10.4} M_{\odot}$ &                       
$6.45\pm1.44$ & $22.10\pm3.06$ & $0.20\pm0.05$ & $-0.69\pm0.31$  \\
                      & $10^{10.4}<M_{*}<10^{11} M_{\odot}$ &
$4.95\pm1.11$ & $27.61\pm4.07$ & $0.19\pm0.09$ & $-0.68\pm0.64$  \\
                      \hline

\multirow{3}{*}{CIV}  & $10^{9.3}<M_{*}<10^{10} M_{\odot}$  & 
$1.03\pm0.33$ & $80.63\pm44.07$ & $0.11\pm0.07$ & 0 (fixed) \\
                      & $10^{10}<M_{*}<10^{10.4} M_{\odot}$ &                       
$2.07\pm0.48$ & $45.96\pm10.29$ & $0.15\pm0.03$ & 0 (fixed) \\
                      & $10^{10.4}<M_{*}<10^{11} M_{\odot}$ & 
                      
$1.31\pm0.09$ & $105.58\pm9.28$ & $0.00\pm0.00$ & 0 (fixed) \\
                      \hline
\end{tabular}
\label{table:rew_mgii_civ}
\end{table*}

\begin{table}[]
\center
\caption{
Best-fit parameters for $W_{0}$ ($r_p/r_{vir}$)
}
\begin{tabular}{l|llll}
    & $A_{w,vir}$             & $r_{w,vir}$   & $B_{w,vir}$              &     $\beta_{w,vir}$     \\
                      \hline
MgII & $5.70\pm0.72$ & $0.19\pm0.02$ & $0.14\pm0.02$ & $-0.67\pm0.27$  \\
CIV & $1.33\pm0.09$ & $0.64\pm0.10$ & $0.07\pm0.05$ & 0 (fixed) \\   
 \hline
\end{tabular}
\label{table:rew_mgii_civ_r_vir}
\end{table}

The results show a number of trends:
\begin{itemize}
    \item \textbf{MgII:} The mean MgII absorption strength correlates with the stellar mass of ELGs. 
    At small impact parameters $r_{p}<30 \, \rm kpc$, the MgII absorption strengths of galaxies with three stellar mass bins are consistent with each other. At larger impact parameters, however, ELGs with higher stellar mass have on average stronger absorption at a given impact parameter. 
    To further describe the mean absorption profiles, we use an empirical functional form with a combination of a power-law function and an exponential function to fit the mean MgII absorption profiles:
\begin{equation}
\label{eq:W}
    \langle W_{0} \rangle = A_{w}\times e^{\frac{-r_{p}}{r_{w}}}+B_{w}\times \bigg( \frac{r_{p}}{\rm 100 \, kpc} \bigg)^{\beta_{w}},
\end{equation}
where, $A_{w}$, $r_{w}$, are the amplitude and characteristic scale of the exponential profile and $B_{w}$, and $\beta_{w}$ are the amplitude and the power of the power-law function. The best-fit curves are shown by the color dashed lines and the best-fit values are summarized in Table~\ref{table:rew_mgii_civ}. 

We also show MgII measurements from the literature at different redshifts. The grey dotted and dashed lines are the best-fit curves from \citet{Huang21} and \citet{Nielsen13} at $z\sim0.4$.
The grey dash-dotted line is the best-fit curve from \citet{Cherrey2025} at $z\sim0.9$.
The green and purple square data points are expected MgII absorption based on CII $\lambda1334$ line at higher redshifts from \citet{Steidel10} and \citet{QPQ14}. We use the median relation between MgII$\lambda2796$ and CII $\lambda1334$ from \citet{lan17} to estimate the MgII absorption strengths. As can be seen, the mean MgII absorption strength around ELGs at $z\sim1.5$ is similar to the level of the absorption strength at $z\sim2$
and is significantly higher than the absorption strength at $z\sim0.4$, indicating a redshift evolution of the MgII gas around galaxies. 

\item{\textbf{CIV:}} The results of CIV absorption are shown in the upper right panel of Figure~\ref{fig:MgII_CIV_REW_mass_dependence}. The mean CIV absorption profiles are more extended than MgII absorption profiles, indicating that the CIV gas and MgII gas have distinct spatial distribution around ELGs. In addition, there is a mild stellar mass dependence of CIV distribution. 
We also use Equation~\ref{eq:W} to fit the mean CIV absorption profiles with the best-fit curves shown by the color dashed lines and parameters listed in Table~\ref{table:rew_mgii_civ}. Similar CIV measurements at $z<0.1$ are shown by the grey data points \citep{Liang14} and grey dashed line \citep{Bordoloi14} and at $z>2$ are shown by the green \citep{Steidel10} and purple data points \citep{QPQ14}. Similarly to the results of MgII, CIV absorption evolves with redshift. 
\end{itemize}

The stellar mass of galaxies correlates with the mass of the dark matter halos in which the galaxies reside \citep[e.g.,][]{wechsler18}. Therefore, we now consider the gas distributions with respect to the virial radius of the dark matter halos ($r_{p}/r_{vir}$).
We use median stellar masses of the three stellar mass bins to estimate the virial radii of the dark matter halos. 
The physical impact parameters are then normalized by the corresponding virial radii. The lower panels of Figure~\ref{fig:MgII_CIV_REW_mass_dependence} show the results. One can see that in this $r_{p}/r_{vir}$ space, the difference of the MgII and CIV absorption profiles around ELGs with different stellar mass is greatly
reduced. The mean absorption profiles around ELGs with different stellar mass now align with each other. 
We use the same functional form (Equation~\ref{eq:W}) to fit the mean absorption profiles:
\begin{equation}
\label{eq:W2}
    \langle W_{0} \rangle = A_{w,vir}\times e^{\frac{-r_{p}/r_{vir}}{r_{w,vir}}}+B_{w,vir}\times \bigg( \frac{r_{p}}{r_{vir}} \bigg)^{\beta_{w,vir}},
\end{equation}
where, $A_{w,vir}$, $r_{w, vir}$ are the amplitude and characteristic scale of the exponential profile and $B_{w,vir}$, and $\beta_{w,vir}$ are the amplitude and the power of the power-law function. The best-fit curves are shown by the black solid lines and the best-fit values are summarized in Table~\ref{table:rew_mgii_civ_r_vir}.
We also perform stacking analysis to cross-validate our measurements. To this end, we obtain composite spectra of background quasars as a function of $r_{p}/r_{vir}$ at the rest-frame of ELGs using the mean estimator. From the high signal-to-noise composite spectra, we measure mean MgII and CIV absorption line strengths as shown in the lower panels of Figure~\ref{fig:MgII_CIV_REW_mass_dependence}, consistent with the measurements based on individual sightlines. We also detect and measure the mean absorption strengths of MgI, FeII, AlII, and AlIII as a function of $r_{p}/r_{vir}$. The details of the analyses and the measurements of other lines are summarized in Appendix~\ref{apen:composite}.

These results indicate that while gas traced by MgII and CIV has different distributions in the halos, their distributions both primarily link to the sizes of the dark matter halos, i.e. gravitational potential of the system. This trend has been reported in previous studies of cool gas traced by MgII and CaII around star-forming galaxies at lower redshifts
\citep[e.g.,][]{Churchill13a, Churchill13b, zhu13b, lan2020, ng25} and our measurements show that the warm gas traced by CIV at $z \sim 1.5$ follows a similar scaling with halo size.

\begin{figure*}
\center
\includegraphics[width=1\textwidth]{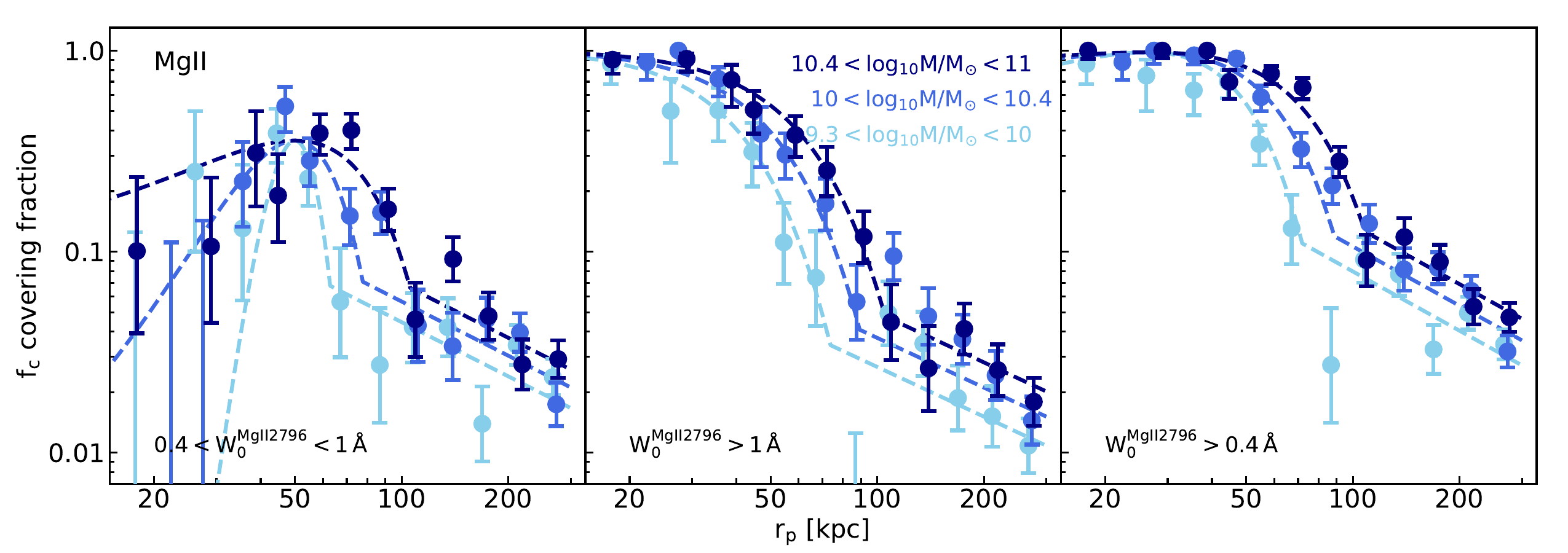}
\includegraphics[width=1\textwidth]{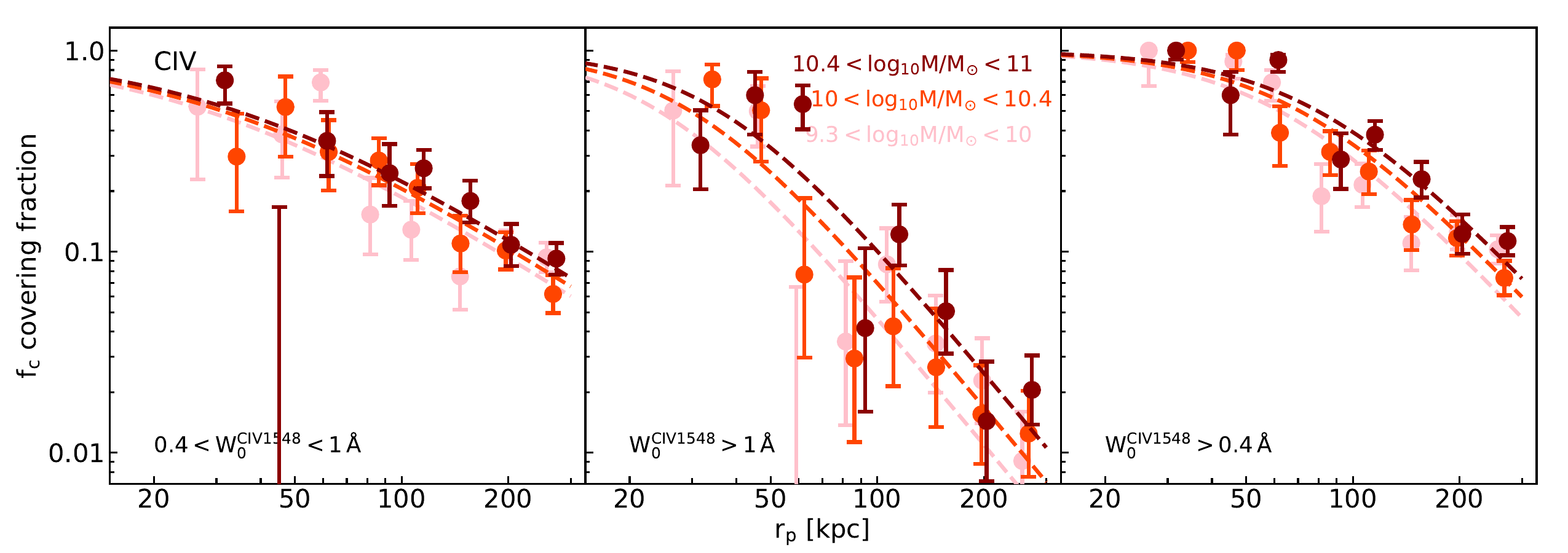}
\caption{Covering fraction of MgII and CIV absorption lines as a function of stellar mass. The upper and lower panels are the measurements of MgII and CIV absorbers respectively. The left, middle and right panels show the results for weak absorbers with $\rm 0.4<W_{0}<1 \, \AA$, strong absorbers with $\rm W_{0}>1 \, \AA$, and all systems with $\rm W_{0}>0.4 \, \AA$ respectively. The stellar mass of galaxies is indicated by the colors. The dashed lines are the best-fit curves with functional forms described in the text. The uncertainties are estimated based on binomial statistics. 
}
\label{fig:MgII_CIV_mass_dependence}
\end{figure*}

\textbf{Covering fraction:}
We now measure the covering fraction of MgII and CIV absorbers around ELGs as a function of impact parameters and stellar mass of galaxies. 
We calculate the covering fraction \citep[e.g.,][]{Bordoloi14} with
\begin{equation}
    f_{c}=\frac{\sum_{i}^{Nabs} w_{i}}{N_{total}},
\end{equation}
where $w_{i}$ is the weight for absorber $i$, $N_{abs}$ is the total number of absorbers, and $N_{\rm total}$ is the total number of sightlines used for searching absorbers. 
We note that because we use the weight to correct for incompleteness, the estimated $f_{c}$ occasionally exceed 1 statistically. In this case, we assign the $f_{c}$ being equal to 1. We note that for those cases, the original $f_{c}$ values are consistent with 1.
We use the Wilson-score interval to estimate the uncertainty of $f_c$. 

\textbf{Gas distribution in physical space:}
Figure~\ref{fig:MgII_CIV_mass_dependence} shows the covering fraction of MgII absorbers in the upper panels and the covering fraction of CIV absorbers in the lower panels. The left panels show the results for both MgII$\lambda2796$ and CIV$\lambda1548$ with $0.4<W_{0}<1 \, \rm \AA$, the middle panels show the results with $W_{0}>1 \, \rm \AA$, and the right panels show the results with $W_{0}>0.4 \, \rm \AA$. The color indicates the $f_{c}$ measurements around galaxies with different stellar mass bins. 
Separating absorbers with $0.4<W_{0}<1 \, \rm \AA$ (weak) and $W_{0}>1 \, \rm \AA$ (strong) is motivated by the fact that the global incidence rate of weak and strong MgII absorbers evolve differently with the incidence rate of strong absorbers following the star-formation density of the Universe  \citep[e.g.,][]{zhu13,matejek2012}. Their distributions around galaxies can be better illustrated with differential measurements. For consistency, we apply the same selections to CIV absorbers.

We now first discuss the properties of MgII absorbers around DESI ELGs. Similar to the rest equivalent width behavior, there is a correlation between the MgII covering fraction and the stellar mass of galaxies for both weak ($0.4<W_{0}<1 \, \rm \AA$) and strong absorbers ($W_{0}>1 \, \rm \AA$), showing that with a fixed impact parameter, the MgII covering fraction increases with stellar mass of galaxies. In addition, the covering fraction of strong and weak MgII absorbers shows different behaviors. The covering fraction of strong absorbers decreases monotonically with increasing impact parameters, while the covering fraction of weak absorbers first increases with impact parameters, reaches the maximum covering fraction at $\sim 40$ kpc for low mass galaxies and $\sim 70$ kpc for galaxies with higher stellar mass, and then decreases with increasing impact parameters.

\begin{table*}[]
\center
\caption{Global best-fit  parameters for $f_{c}^{MgII}$ as a function of $W_{0}^{\lambda2796}$ in $r_{p}$ space}
\begin{tabular}{c|ccccccc}
                   & $A_{c}$ & $r_{c}$ [kpc] & $E_{c}$ [kpc] & $\alpha_{c}$ & $B_{c}$ & $\beta_c$ & $\gamma_c$ \\
                    \hline
$0.4<W_{0}<1 \rm \, \AA$ & $0.36\pm0.07$ & $49.96\pm3.25$ & $10.70\pm3.79$ & $0.75\pm0.21$ & $0.05\pm0.01$ & $-0.90\pm0.33$ & $0.23\pm0.14$\\
\hline
$W_{0}>1 \rm \, \AA$ & $1.00\pm0.00$ & $3.98\pm5.84$ & $30.30\pm4.40$ & $0.22\pm0.06$ & $0.032\pm0.005$ & $-0.82\pm0.24$ & $0.32\pm0.13$ \\
\hline
$W_{0}>0.4 \rm \, \AA$ & $0.98\pm0.01$ & $26.21\pm4.27$ & $26.04\pm3.71$ & $0.32\pm0.08$ & $0.09\pm0.02$ & $-0.98\pm0.25$ & $0.28\pm0.09$ \\
\hline
\end{tabular}
\label{table:fc_mgii}
\end{table*}

\begin{table}[]
\center
\caption{
Global best-fit parameters for $f_{c}^{CIV}$ as a function of $W_{0}^{\lambda1548}$ in $r_{p}$ space
}
\begin{tabular}{c|ccc}
 & $C$ & $D$ & $\delta_{c}$ \\
                    \hline
$0.4<W_{0}<1 \rm \, \AA$ & $2.68\pm0.51$ & $1.47\pm0.15$ & $0.029\pm0.038$  \\
\hline
$W_{0}>1 \rm \, \AA$ & $4.91\pm1.31$ & $1.43\pm0.15$ & $0.060\pm0.045$ \\
\hline
$W_{0}>0.4 \rm \, \AA$ & $4.40\pm0.45$ & $1.82\pm0.04$ & $0.030\pm0.015$ \\
\hline

\end{tabular}
\label{table:fc_civ}
\end{table}

To capture the covering fraction of strong and weak MgII absorbers, we use an empirical piecewise functional form combining a Gaussian distribution capturing the inner covering fraction and a power law capturing the outer covering fraction:
\begin{equation}
f_{c}^{\rm MgII} (r_{p}) = 
\begin{dcases}
    A_{c}\times e^\frac{-(r_p-r_c)^{2}}{2\sigma_{c}^{2}}, & \text{when} \,r_{p}\leq r_{b}\\
    B_{c}\times\bigg(\frac{M_{*}}{10^{10} M_\odot}\bigg)^{\gamma_{c}}\bigg(\frac{r_p}{100 \, \rm kpc}\bigg)^{\beta_{c}},              &  \text{when} \,r_{p} > r_{b},
\end{dcases}
\label{eq_1_MgII}
\end{equation}
where $\sigma_c$ is the width parameter of the Gaussian profile
\begin{equation}
    \sigma_{c}=E_{c}\times\bigg(\frac{M_{*}}{10^{10} M_\odot}\bigg)^{\alpha_{c}},
\end{equation}
with amplitude $E_{c}$ and stellar mass dependence parameter $\alpha_{c}$.
$A_{c}$ and $r_{c}$, are the best-fit amplitude, center location of the Gaussian profile. $B_{c}$ is the best-fit covering fraction at 100 kpc for $10^{10} \, M_{\odot}$ galaxies, $\gamma_{c}$ is the stellar mass dependence, and $\beta_{c}$ is the slope of the power law. $r_b$ is the impact parameter at the transition of the two functional forms which is the location where the two functional forms produce the same value of the covering fraction. 
The fitting is performed globally with a single set of parameters for the three stellar mass bins in each $W_{0}$ selection.
The dashed lines shown in the upper panels of Figure~\ref{fig:MgII_CIV_mass_dependence} are the best-fit profiles with Equation~\ref{eq_1_MgII}, adequately capturing the distributions of strong and weak MgII absorbers. The best-fit parameters are summarized in Table~\ref{table:fc_mgii}. 
We perform the least-square regression fitting which assumes symmetric uncertainties. We adopt the larger uncertainty value from the Wilson score interval and use the uncertainty as the weight for each data point. The uncertainties of the best-fit parameters are obtained via bootstrapping the sample 500 times. As shown by the best-fit curves, the $r_{b}$ values increase with stellar mass, indicating that the extension of the inner gas distribution correlates with stellar mass. 
Based on the best-fit parameter $\alpha_{c}$ ($0.32\pm0.08$) for the stellar mass dependence, the stellar mass dependence is detected with $4 \sigma$ for the $W_{0}>0.4 \rm \, \AA$ bin.
The outer power law also shows stellar mass dependence, $f_{c}\propto M_{*}^{\gamma_{c}}$, with $\gamma_{c}=0.28\pm0.09$ for $W_{0}>0.4 \rm \, \AA$ bin. This trend is consistent with previous measurements of low-ionized absorption species, including HI, MgII and CaII \citep[e.g.,][]{Bordoloi18,  Chen2025, Cherrey2025, ng25}.

For CIV absorbers, their covering fraction is shown in the lower panels of Figure~\ref{fig:MgII_CIV_mass_dependence}. The covering fraction of both weak and strong CIV absorbers decreases with increasing impact parameters. The covering fraction of strong CIV absorbers declines with impact parameters faster than the covering fraction of weak CIV absorbers. One can also observe the trend that the covering fraction of CIV absorbers around galaxies with higher mass tends to be higher than the covering fraction around galaxies with lower mass. However, the difference in covering fraction of CIV absorbers is smaller than that of MgII absorbers. 

To summarize the covering fraction of CIV absorbers, we use a functional form adopted from \citet{Schroetter21}:

\begin{equation}
f_{c}^{\rm CIV}(r_{p}) = \frac{1}{1+e^{t}} \, ,
\label{eq_1_CIV}
\end{equation}
\begin{equation}
t = C\times\bigg[\rm log_{10}\frac{r_{p}}{kpc}-D\times\bigg(\frac{M_{*}}{10^{10} M_{\odot}}\bigg)^{\delta_{c}}\bigg] \, ,
\label{eq_2_CIV}
\end{equation}
where C, D, and $\delta_{c}$ are the best-fit free parameters. 
The fitting is performed globally with a single set of parameters for three stellar mass bins in each $W_{0}$ selection.
The dashed lines shown in the lower panels of Figure~\ref{fig:MgII_CIV_mass_dependence} are the best-fit profiles with Equation~\ref{eq_1_CIV}. The best-fit parameters are summarized in Table~\ref{table:fc_civ}. 
Based on the best-fit profiles, one can observe a mild stellar mass dependence of CIV $f_{c}$ with $2 \sigma$ for the $W_{0}>0.4 \rm \, \AA$ bin based on the best-fit parameter value $\delta_{c}$ ($0.030\pm0.015$). This tentative trend is consistent with previous findings of a higher CIV detection rate around more massive galaxies at $z\sim0$ \citep{burchett16} and $z<2$ \citep{Dutta21}.

The $f_{c}$ of MgII and CIV shows different behaviors similar to their rest equivalent width behavior. In the inner halos with $r_{p}<50$ kpc, the $f_c$ of MgII and CIV absorbers with $W_{0}>0.4\,\rm \AA$ are nearly $100\%$. However, beyond 50 kpc, the $f_{c}$ of MgII absorbers decreases faster than the $f_{c}$ of CIV absorbers as a function of $r_{p}$. For example, at 100 kpc, for galaxies with $M_{*}>10^{10.4} M_{\odot}$, the $f_{c}$ of CIV is 2 times higher than the $f_{c}$ of MgII, mostly contributed by the $f_{c}$ of weak absorbers with $0.4<W_{0}<1 \rm\, \AA$ as shown in the right panels of Figure~\ref{fig:MgII_CIV_mass_dependence}. 
On the other hand, strong absorbers of MgII and CIV have similar $f_{c}$ from the inner regions to the outer regions of the halos. The difference of the $f_{c}$ of MgII and CIV absorbers demonstrates that the warmer gas traced by CIV absorption lines covers larger spatial regions around galaxies than the cool gas traced by MgII absorption lines.

\begin{figure*}
\center
\includegraphics[width=1\textwidth]{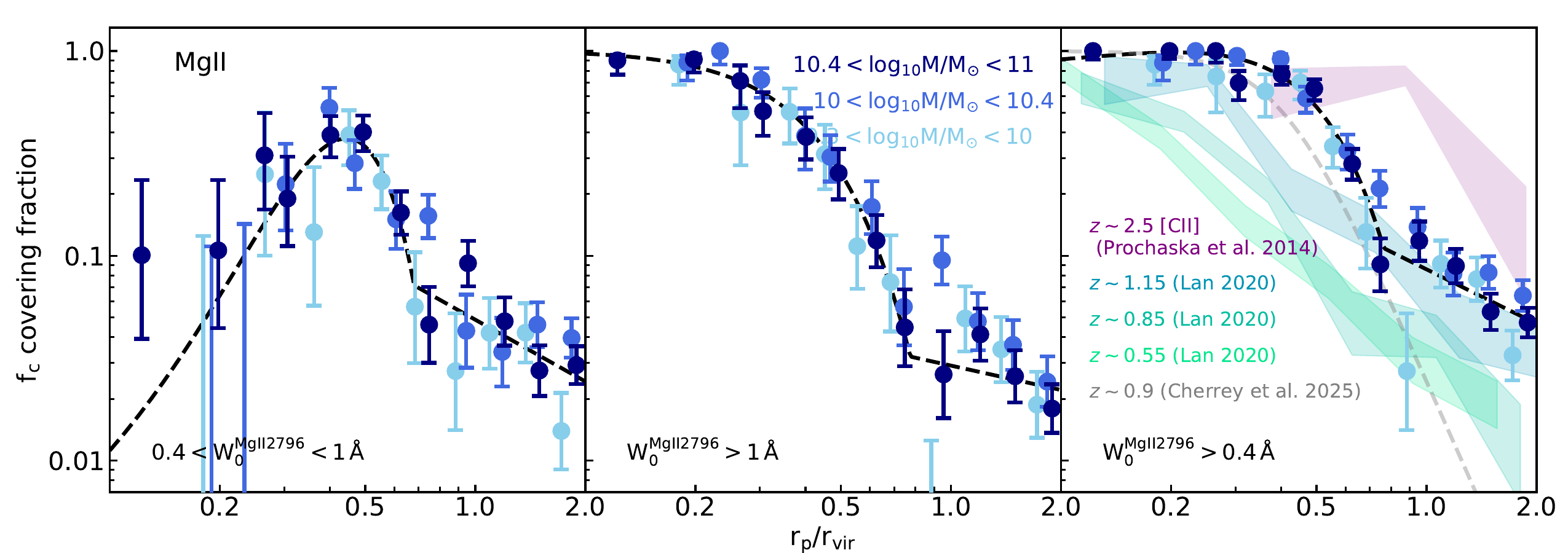}
\includegraphics[width=1\textwidth]{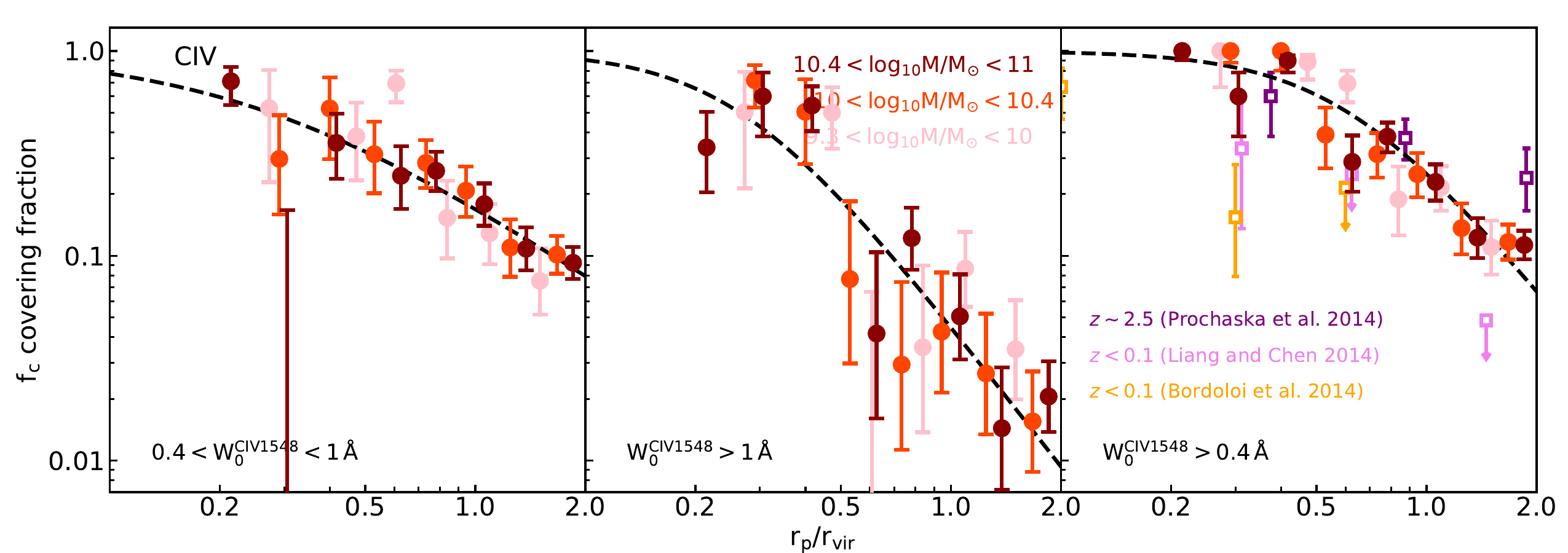}
\caption{
Covering fraction of MgII and CIV absorption lines as a function of $r_{p}/r_{vir}$. The upper and lower panels are the measurements of MgII and CIV absorbers respectively. The left, middle and right panels show the results for weak absorbers with $\rm 0.4<W_{0}<1 \, \AA$, strong absorbers with $\rm W_{0}>1 \, \AA$, and all systems with $\rm W_{0}>0.4 \, \AA$ respectively. The stellar mass of galaxies is indicated by the colors. The dashed lines are the best-fit curves with functional forms described in the text. The uncertainties are estimated based on binomial statistics. In the upper right panel, the green bands are $f_{c}$ measurements around star-forming galaxies with $10^{10.5} M_{\odot}$ from \citet{lan2020} ($z<1.2$), the purple band is $f_{c}$ measurements with CII from \citet{QPQ14} ($z\sim2.5$), and the grey dashed line is the best fit MgII $f_{c}$ from \citet{Cherrey2025}. In the lower right panel, the orange and pink square data points are measurements based on data from \citet{Bordoloi14} and \citet{Liang14} at $z<0.1$. Data points with arrows pointing down are $3\sigma$ upper limits. The purple data points are measurements based on data from \citet{QPQ14} at $z\sim2.5$.
} 
\label{fig:MgII_CIV_rvir}
\end{figure*}

\begin{figure*}
\center
\includegraphics[width=1\textwidth]{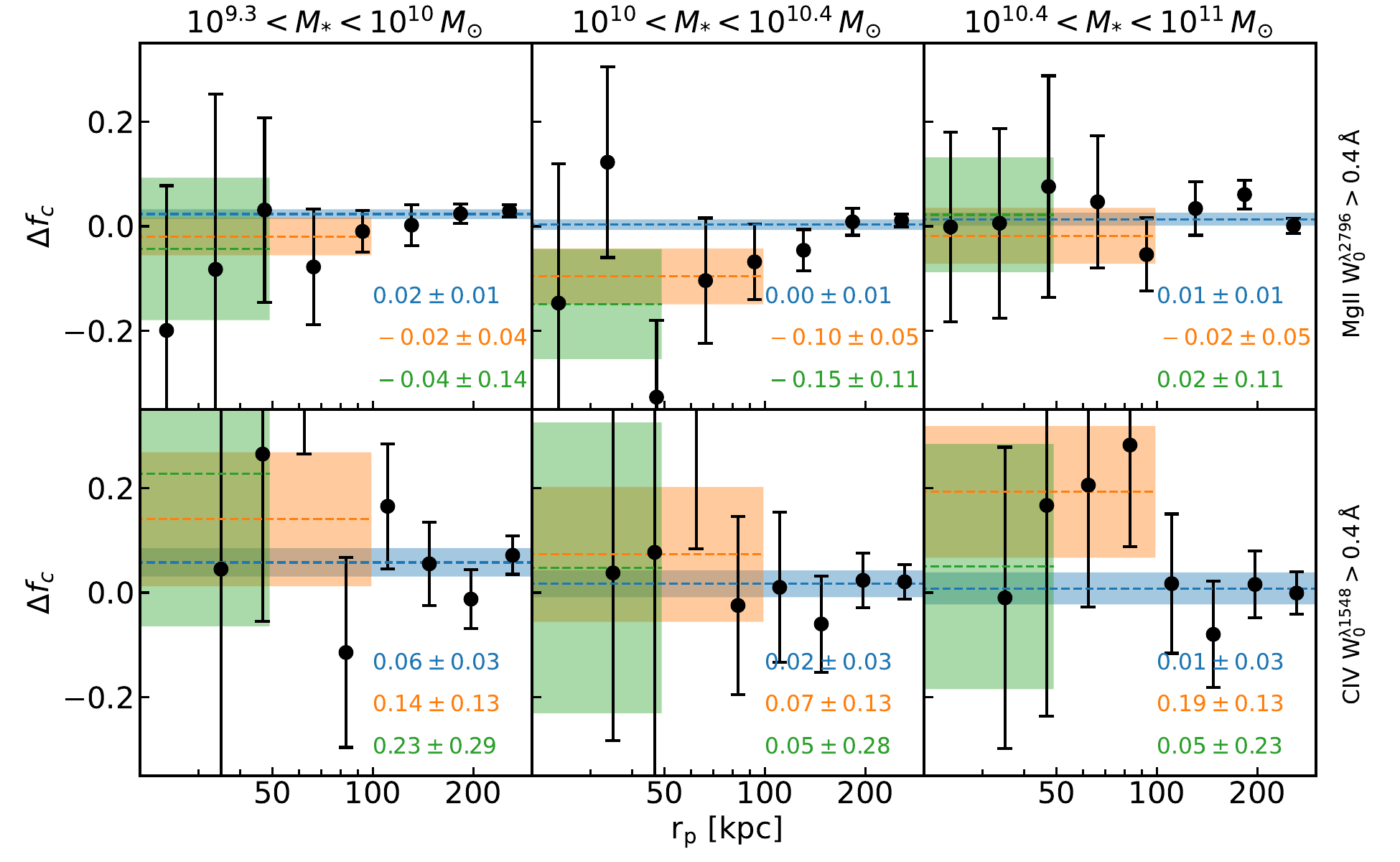}
\caption{Covering fraction difference (covering fraction of high-SFR galaxies - covering fraction of low-SFR galaxies). 
The left, middle, and right panels show the covering fraction differences for galaxies with $10^{9.3}<M_{*}<10^{10} \, M_{\odot}$, $10^{10}<M_{*}<10^{10.4} \, M_{\odot}$ and $M_{*}>10^{10.4}\, M_{\odot}$ respectively. The upper and lower panels show the results for MgII absorbers and CIV absorbers with $W_{0}>0.4 \rm \, \AA$ respectively. The color bands show the best-fit values from inner $r_{p}<50 \, \rm kpc$ (green), intermediate $r_{p}<100 \, \rm kpc$ (orange), to outer regions $r_{p}<300 \, \rm kpc$ (blue).
}
\label{fig:MgII_CIV_SFR}
\end{figure*}

\textbf{Gas distribution with respect to the virial radius:}
We now consider the gas distribution with respect to the virial radius.
The $f_{c}$ measurements with $r_{p}/r_{vir}$ are shown in Figure~\ref{fig:MgII_CIV_rvir}. 
Again, the upper and lower panels show the measurements of MgII and CIV respectively. We find that after normalizing the physical impact parameters by the virial radius of the halos, both the $f_{c}$ of MgII and CIV of galaxies with different mass align with each other. The correlations between $f_{c}$ of MgII and CIV and stellar mass in the physical impact parameter space shown in Figure~\ref{fig:MgII_CIV_mass_dependence} disappear. 

We combine all the measurements $<2\,r_{vir}$ and perform global fitting for each $W_{0}$ selection. For MgII, we use the functional form as Equation~\ref{eq_1_MgII} but change the $r_{p}$ into $r_{p}/r_{vir}$:
\begin{equation}
f_{c}^{\rm MgII} \bigg(\frac{r_{p}}{r_{vir}}\bigg) = 
\begin{dcases}
    A_{c, vir}\times e^\frac{-(\frac{r_p}{r_{vir}}-r_{c,vir})^{2}}{2\sigma_{c,vir}^{2}}, &\text{when} \,\frac{r_{p}}{r_{vir}}\leq r_{b, vir}\\
    B_{c, vir}\times\bigg(\frac{r_p}{r_{vir}}\bigg)^{\beta_{c, vir}},              &\text{when} \,\frac{r_{p}}{r_{vir}} > r_{b, vir},
\end{dcases}
\label{eq_MgII_halo}
\end{equation}
where $A_{c,vir}$, $r_{c,vir}$, and $\sigma_{c,vir}$ are the best-fit amplitude, center location, and the width parameters of the Gaussian profile, and $B_{c,vir}$ is the best-fit covering fraction at $r_{vir}$ and $\beta_{c, vir}$ is the slope of the power law. $r_{b,vir}$ is the impact parameter at the transition of the two functional forms which is the location where the two functional forms produce the same value of the covering fraction. The values are listed in Table~\ref{table:fc_mgii_rvir}.

\begin{table*}[]
\center
\caption{
Global best-fit parameters for $f_{c}^{MgII}$ and $W_{0}^{\lambda2796}$ 
}
\begin{tabular}{c|ccccc}
                   & $A_{c,vir}$ & $r_{c,vir}$ & $\sigma_{c,vir}$ & $B_{c,vir}$ & $\beta_{c,vir}$ \\
                    \hline
$0.4<W_{0}<1 \rm \, \AA$ & $0.37\pm0.09$ & $0.44\pm0.03$ & $0.13\pm0.04$ & $0.05\pm0.01$ & $-1.00\pm0.63$ \\
$W_{0}>1 \rm \, \AA$ & $1.00\pm0.00$ & $0.03\pm0.05$ & $0.28\pm0.03$ & $0.03\pm0.005$ & $-0.40\pm0.36$ \\
$W_{0}>0.4 \rm \, \AA$ & $0.98\pm0.01$ & $0.21\pm0.03$ & $0.27\pm0.02$ & $0.09\pm0.01$ & $-0.88\pm0.37$ \\
\hline
\end{tabular}
\label{table:fc_mgii_rvir}
\end{table*}

\begin{table}[]
\center
\caption{
Global best-fit parameters for $f_{c}^{CIV}$ and $W_{0}^{\lambda1548}$ 
}
\begin{tabular}{c|cc}
                   & $C_{vir}$ & $D_{vir}$ \\
                    \hline
$0.4<W_{0}<1 \rm \, \AA$ & $2.84\pm0.54$ & $-0.56\pm0.13$\\
$W_{0}>1 \rm \, \AA$ & $5.31\pm1.56$ & $-0.58\pm0.13$ \\
$W_{0}>0.4 \rm \, \AA$ & $5.12\pm0.61$ & $-0.21\pm0.03$ \\
\hline
\end{tabular}
\label{table:fc_civ_rvir}
\end{table}

Similarly, for CIV, we use the functional form as Equation~\ref{eq_1_CIV}
and change the $r_{p}$ into $r_{p}/r_{vir}$:
\begin{equation}
f_{c}^{\rm CIV}\bigg(\frac{r_{p}}{r_{vir}}\bigg) = \frac{1}{1+e^{t_{vir}}} \, ,
\label{eq_1_CIV_halo}
\end{equation}
\begin{equation}
t_{vir} = C_{vir}\times\bigg(\rm log_{10}\frac{r_{p}}{r_{vir}}-D_{vir}\bigg) \, ,
\label{eq_2_CIV}
\end{equation}
where $C_{vir}$ and $D_{vir}$ are the best-fit free parameters.
The best-fit curves are shown by the black dashed lines in the panels and the best-fit parameter values are listed in Table~\ref{table:fc_civ_rvir}.
The results are consistent with the rest equivalent width measurements discussed previously, indicating that the mechanisms regulating the halo gas distribution are closely linked to the size of dark matter halos.

\textbf{Redshift trend:}
In the right panels of Figure~\ref{fig:MgII_CIV_rvir}, we also show MgII and CIV covering fractions at different redshifts from previous studies. In the top right panels, the color bands show the measurements with redshifts listed in the figure from \citet{lan2020} and from \citet{QPQ14} based on CII absorption lines. The grey dashed line shows the best-fit profile from \citet{Cherrey2025} at $z\sim0.9$. 
We note that the gas properties from \citet{QPQ14} are measured around quasars instead of star-forming galaxies. As can be seen, the MgII covering fraction measured within the dark matter halos increases with redshifts and our new measurements at $z\sim1.5$ continue the trend observed around star-forming galaxies at lower redshifts. 

For CIV absorbers, we compare the covering fractions with $W_{0}^{\lambda1548}>0.4 \rm \, \AA$ from this work with measurements at the local Universe and at $z\sim2$ in the lower right panel of Figure~\ref{fig:MgII_CIV_rvir}. The orange and pink square data points are the covering fraction estimation based on data at $z<0.1$ from \citet{Bordoloi14} and \citet{Liang14} respectively. The purple square data points are from \citet{QPQ14} measured around quasars at $z\sim2$. At $z<0.1$, there is no detection for CIV with $>0.4 \rm\, \AA$ at $r_{p}>0.5 \, r_{vir}$ shown by the 3$\sigma$ upper limits of the square data points. On the other hand, the covering fractions are $\sim0.4$ around $z\sim1.5$ galaxies and $z\sim2$ quasars, indicating a redshift evolution of CIV gas distribution in the halos. This result is consistent with the result from \citet{Dutta21} who probed MgII and CIV absorption mostly beyond the virial radius of the halos of galaxies with stellar mass and SFR down to $\sim5\times 10^{8} \,M_{\odot}$ and $0.1\, M_{\odot} \,yr^{-1}$. 

We note that \citet{Schroetter21} also investigated the redshift evolution of MgII and CIV covering fraction around star-forming galaxies at $1<z<1.5$. They reached to the same conclusion for MgII absorption, while found that the CIV covering fraction tends to increase toward lower redshifts, a reverse trend as shown in Figure~\ref{fig:MgII_CIV_rvir}. This inconsistency might be due to that in \citet{Schroetter21}, they detected and included weaker CIV absorbers with $0.1<W_{0}<0.4 \, \rm \AA$ in their $f_{c}$ calculation. In \citet{Schroetter21}, approximately $30\%$ of their detected sources have $0.1<W_{0}<0.4\, \rm \AA$ and in \citet{Bordoloi14}, approximately $50\%$ of sources have $0.1<W_{0}<0.4 \, \rm \AA$. Such a population which might evolve differently in comparison with stronger absorption systems is not included in our analysis. 

\begin{figure}
\center
\includegraphics[width=0.45\textwidth]{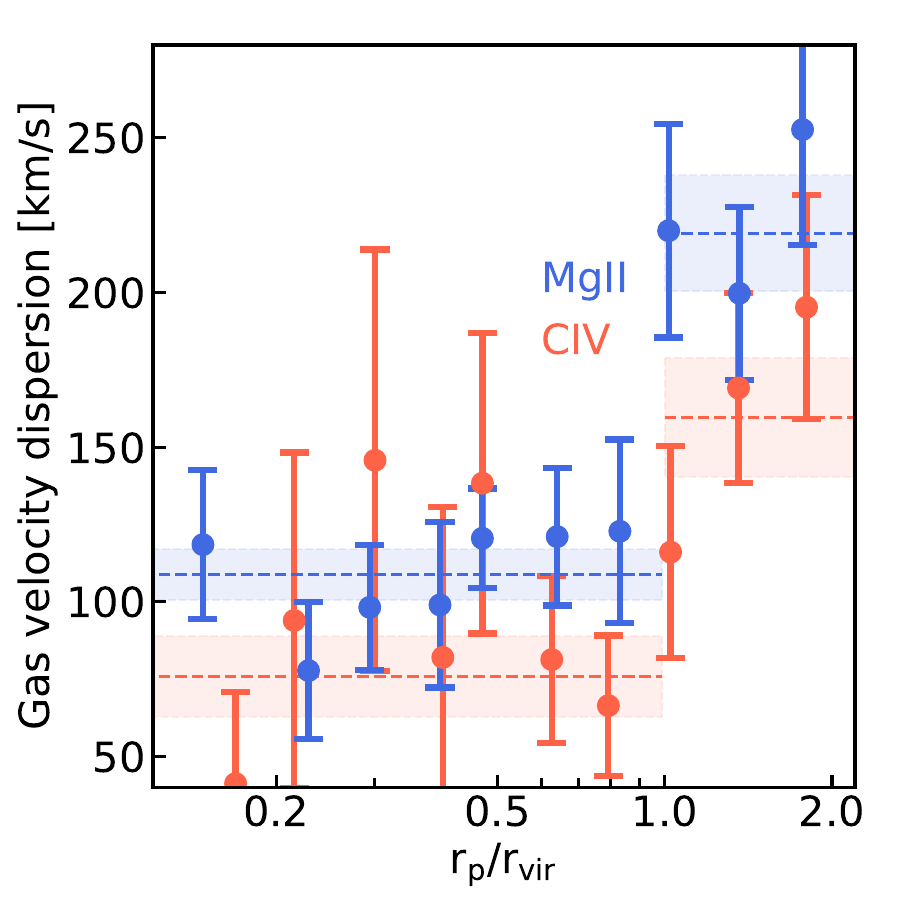}
\caption{Gas velocity dispersion as a function of $r_{p}/r_{vir}$. The blue and red data points show the measurements of MgII and CIV. The dashed lines and shaded regions show the best-fit gas velocity dispersion values with $r_{p}<r_{vir}$ and $r_{p}>r_{vir}$ respectively. The uncertainties of the data points are estimated via bootstrapping the sample 500 times.} 
\label{fig:MgII_CIV_velocity_dispersion}
\end{figure}

\begin{figure*}
\center
\includegraphics[width=0.8\textwidth]{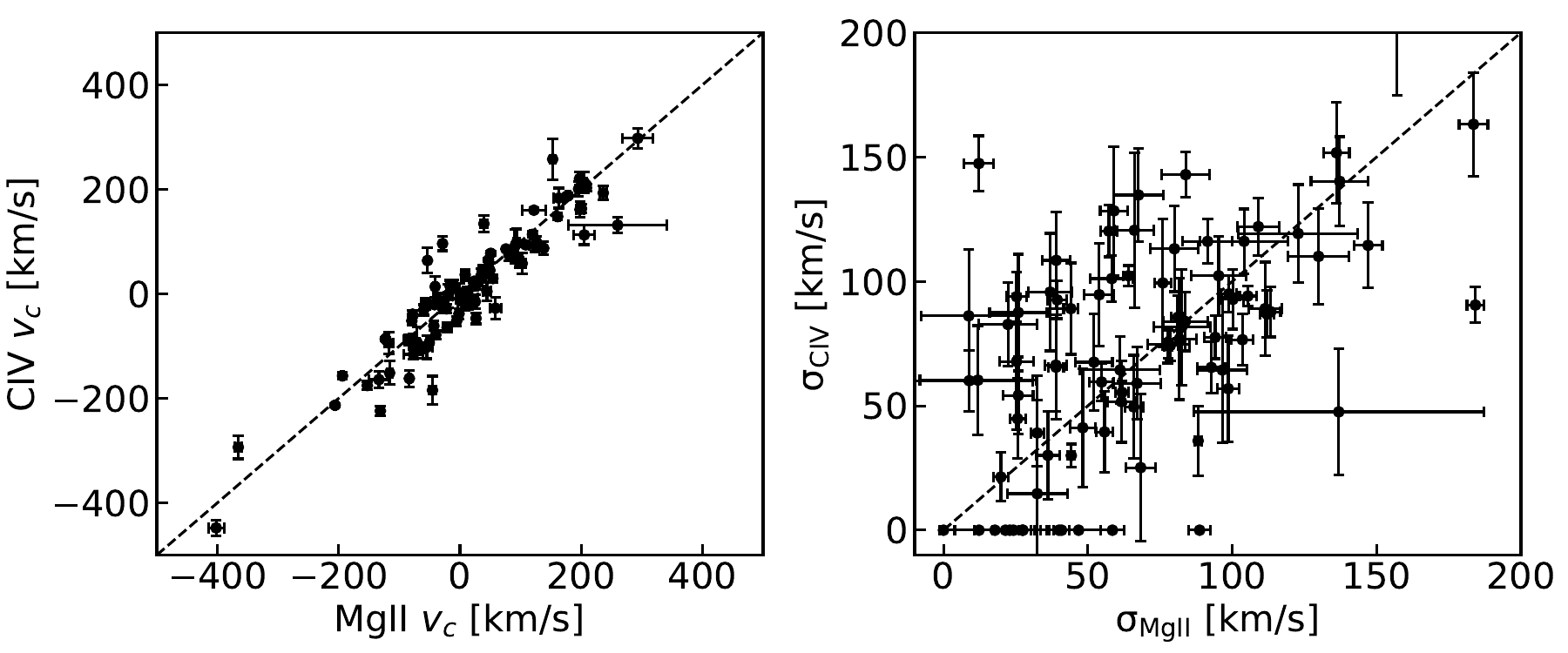
}
\caption{MgII and CIV interconnection. \emph{Left}: the central velocities of MgII and CIV with respect to the ELGs. \emph{Right:} the intrinsic line widths of MgII ($\sigma_{MgII}$) and CIV ($\sigma_{CIV}$). We note that there are $\sim1\% \, (1/87)$ of systems with the best-fit $\sigma_{MgII}=0 \rm \, km/s$ and $\sim15\% \, (13/87)$ of systems with the best-fit $\sigma_{CIV}=0 \rm \, km/s$. Those systems have intrinsic line widths significantly smaller than the DESI spectral resolution ($\sigma_{resolution}\sim34 \, \rm km/s$ for MgII and $\sim62 \, \rm km/s$ for CIV).
 } 
\label{fig:MgII_CIV_dv}
\end{figure*}

\subsubsection{SFR dependence}
We now explore the correlation between the gas distributions and SFR of galaxies. Given that the SFR correlates with the stellar mass of galaxies, we further fix the range of stellar mass and compare the covering fraction of DESI ELGs with SFR higher and lower than the median value of the SFR of the sample. Figure~\ref{fig:MgII_CIV_SFR} shows the difference between the covering fraction around galaxies with higher SFRs and with lower SFRs: $\rm \Delta f_{c} \equiv f_{c}(high\, SFR)-f_{c}(low\,SFR)$. The results for MgII and CIV both with $W_{0}>0.4\, \rm \AA$ are shown in the upper and lower panels. The results of galaxies with different stellar mass cuts are shown from left to right with increasing stellar mass. The green, orange, and blue horizontal dashed lines and shaded regions indicate the best-fit $\Delta f_{c}$ values and the uncertainty ranges, listed in each panel,  when considering $\Delta f_{c}$ measurements within 50 kpc, 100 kpc, and 300 kpc respectively. 

The results show that there is no significant correlation between SFR and $f_{c}$ of both MgII and CIV absorbers. We also perform the same measurements for weak and strong MgII and CIV absorbers separately and do not find any significant correlation. 

This result is inconsistent with some of the previous studies. For example, \citet{anand21} and \citet{lan18} show that MgII $f_{c}$ increases with SFR especially in the inner halos. \citet{ng25} also find that with a fixed stellar mass, the CaII absorption strength correlates with SFR by $\propto SFR^{0.3}$ at $z\sim0.2$. 
We argue that if the CGM properties indeed correlate with SFR at $z\sim1.5$, this inconsistency can be due the dynamic range of SFR covered by DESI ELGs. DESI ELGs are star-forming galaxies with preferentially higher SFRs than that of star-forming main sequence galaxies \citep{lan24}. The largest SFR difference between the high SFR sample (median SFR $\sim40\, M_{\odot}\,yr^{-1}$) and low SFR sample (median SFR $\sim20\, M_{\odot}\,yr^{-1}$) is a factor of 2 from the $M_{*}>10^{10.4} \, M_{\odot}$ selection. This yields only an $\sim20\% \, (2^{0.3})$ difference in absorption strength based on \citet{ng25}. 
This might limit us to detect any correlation between gas properties and SFR. It is also possible that there is no correlation between the CGM properties and SFR intrinsically at $z\sim1.5$. A large sample covering a wider SFR range is required to identify the correlation between gas and SFR if any.

\subsection{Gas kinematics}
\label{text:kinematics}

The kinematics of gas around galaxies traced by MgII and CIV absorption lines is now explored. We measure the difference, $\Delta v$, between the central velocities of galaxies and the central velocities of the absorption line systems along the lines of sight as a function of $r_{p}/r_{vir}$. We calculate the standard deviation of the $\Delta v$ distribution, the gas velocity dispersion, using the median absolute deviation (MAD) estimator. The results are summarized in Figure~\ref{fig:MgII_CIV_velocity_dispersion} with blue data points showing the gas velocity dispersion of MgII absorbers and red data points showing that of CIV absorbers. 
We find that within the virial radius, both the gas velocity dispersions of MgII and CIV are $\sim 100$ $\rm km \, s^{-1}$, consistent with the velocity dispersion of dark matter particles with $\sim10^{12} M_{\odot}$ halo \citep[e.g.,][]{Elahi18} and previous measurements around star-forming galaxies \citep[e.g.,][]{lan18, anand21}. On the other hand, the velocity dispersions increase to $\sim 200$ $\rm km \, s^{-1}$ outside of the halo. The color dashed lines show the best-fit values of the velocity dispersions with $0.1<r_p/r_{vir}<1$ and $1<r_p/r_{vir}<2$ for MgII (blue) and CIV (red) absorption lines, indicating that the velocity dispersions of the cool gas traced by MgII and the warm gas traced by CIV are consistent with each other. The increasing gas velocity dispersions at larger scale can be due to the contribution of the gas associated with nearby galaxies 
\citep[so-called two-halo terms; e.g.,][]{zhu14}. 
We note that the velocity dispersions increase from $\sim 100 \, \rm km \, s^{-1}$  to $\sim 200 \, \rm km \, s^{-1}$ at around $0.8-1 \, r_{vir}$, especially for the gas traced by MgII. This scale aligns with the transition of the MgII covering fraction around galaxies ($\sim0.8 \, r_{vir}$) (the top right panel of Figure~\ref{fig:MgII_CIV_rvir}), suggesting that the bulk of the gas in and beyond this scale might have different origins.

\begin{figure*}
\center
\includegraphics[width=0.8\textwidth]{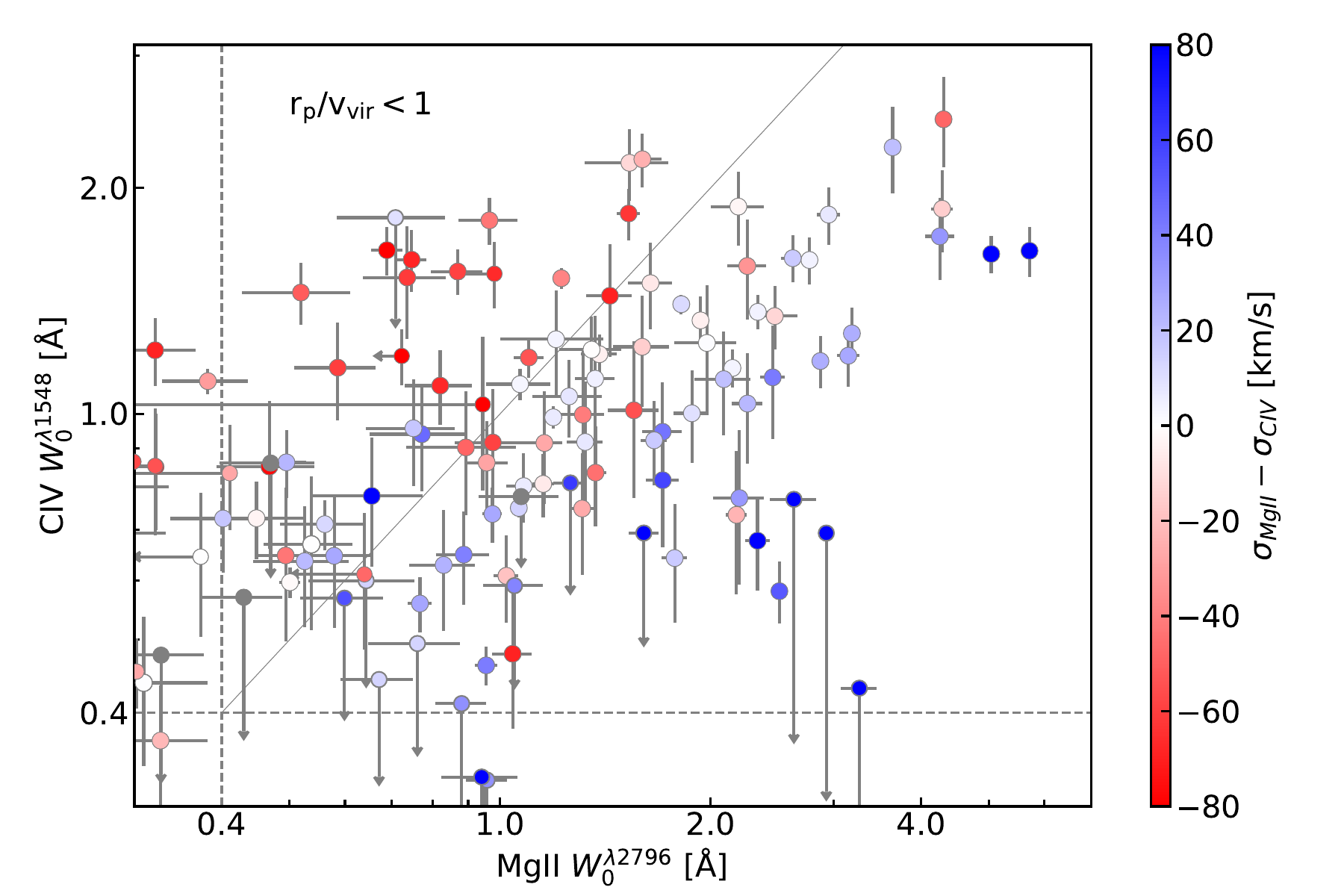}
\caption{MgII $W_{0}^{\lambda2796}$ and CIV $W_{0}^{\lambda1548}$ plane.  The x-axis shows the MgII $W_{0}^{\lambda2796}$ and the y-axis shows the CIV $W_{0}^{\lambda1548}$. The colors indicate the line width differences between MgII and CIV. Blue indicates $\sigma_{MgII}>
\sigma_{CIV}$ and red indicates $\sigma_{MgII}<\sigma_{CIV}$. Systems with MgII detection and without CIV detection are shown with upper limits in y-axis (pointing down). Systems without MgII detection and with CIV detection are shown with upper limits in x-axis (pointing left). For the non-detection, we show the best-fit $W_{0}$ values.} 
\label{fig:MgII_CIV_line_width}
\end{figure*}

\subsection{MgII and CIV connections}
\label{text:MgIIvsCIV}
We now explore how the properties of MgII and CIV absorbers correlate with each other around ELGs. To this end, we focus on the properties of MgII and CIV detected in quasar spectra with (1) $S/N>4$ around both MgII and CIV spectral regions, (2) $W_{0}^{\lambda2796}>0.4 \, \rm \AA$ and $W_{0}^{\lambda1548}>0.4 \, \rm \AA $ and (3) with $r_p <r_{vir}$ around ELGs. 
The left panel of Figure~\ref{fig:MgII_CIV_dv} shows the central velocities of MgII (x-axis) and CIV (y-axis) with respect to the galaxies and the right panel of Figure~\ref{fig:MgII_CIV_dv} shows the best-fit  intrinsic line width, $\sigma$, of MgII (x-axis) and CIV (y-axis) lines. The left panel shows that there are systems with velocity differences larger than the measured uncertainties, reflecting the intrinsic velocity differences of the gas along the sightlines. The line width distribution of MgII and CIV shown in the right panel also demonstrates the intrinsic difference of absorption profiles. For example, with a fixed $\sigma_{\rm MgII}$ being 50 $\rm km \, s^{-1}$, the $\sigma_{\rm CIV}$ ranges from smaller than $\sigma_{\rm MgII}$ to larger than $\sigma_{\rm MgII}$ by a factor of 2-3. This result indicates that the multiple absorption components of the cool and warm gas traced by MgII and CIV along the same sightlines do not have the same velocity distributions. We will discuss the implication of this result in Section~\ref{sec:CGM_structure}. 

To further illustrate the variation of MgII and CIV absorption properties, we show MgII and CIV rest equivalent widths with colors indicating the line width differences in Figure~\ref{fig:MgII_CIV_line_width}. In addition, MgII detections without CIV detections are shown with upper limits in y-axis (arrow pointing down) and CIV detections without MgII detections are shown with upper limits in x-axis (arrow pointing left). For MgII absorbers with $W_{0}^{\lambda 2796}<1 \rm \AA$, with a fixed MgII rest equivalent width, the associated CIV rest equivalent widths range from being smaller than the MgII absorption (below the grey solid line) to being larger than the MgII absorption (above the grey solid line). For systems with stronger CIV absorption than MgII absorption, the CIV line widths are preferentially larger than the MgII line widths indicated by the red colors and vice versa. This indicates that the metal absorption line with stronger absorption is driven by the broader kinematics distribution, which is expected for saturated absorption-line systems \citep[e.g.,][]{Churchill03}. However, for strong MgII absorbers with $W_{0}^{\lambda 2796}>1 \rm \AA$, the majority of systems has $W_{0}^{\lambda 1548}/W_{0}^{\lambda2796}<1$ (below the grey solid line) and the MgII line widths are mostly larger than the CIV line widths. The ratio ($W_{0}^{\lambda 1548}/W_{0}^{\lambda2796}$) becomes even smaller for MgII systems with higher $W_{0}^{\lambda2796}$. This behavior suggests that the mechanisms giving rise to the strong MgII absorbers preferentially produce a
broader kinematic distribution of the cool gas components than the warmer gas components.

\begin{figure*}
\center
\includegraphics[width=0.9\textwidth]{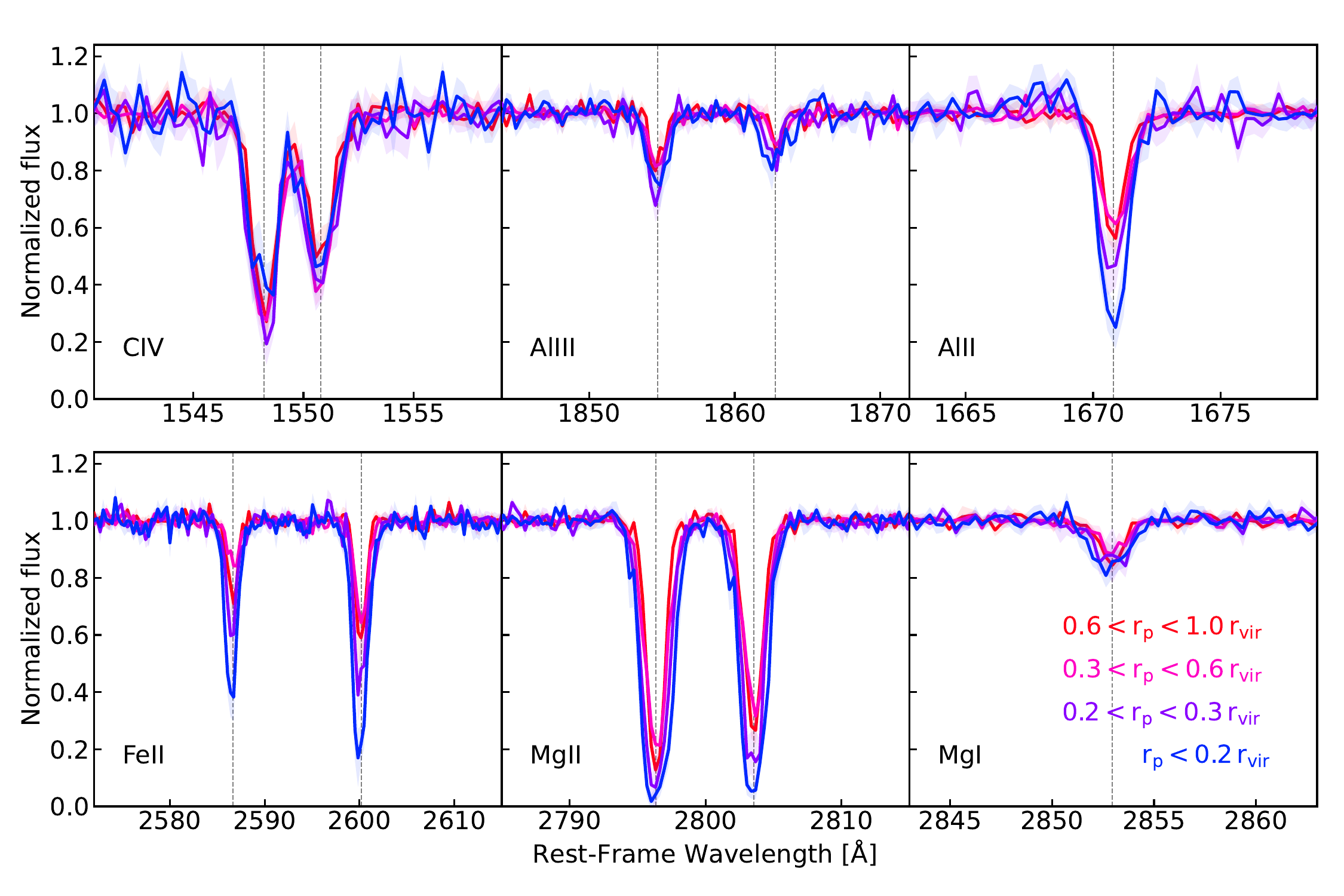}
\caption{Median composite spectra of MgII absorbers with $W_{0}^{\lambda2796}>1 \, \rm \AA$ as a function of $r_{p}/r_{vir}$. The colors indicate the $r_{p}/r_{vir}$ bins from blue (inner region) to red (outer region). The shaded regions show the spectral uncertainty based on bootstrapping the sample 500 times.} 
\label{fig:composite_spectra}
\end{figure*}

\begin{figure}
\center
\includegraphics[width=0.45\textwidth]{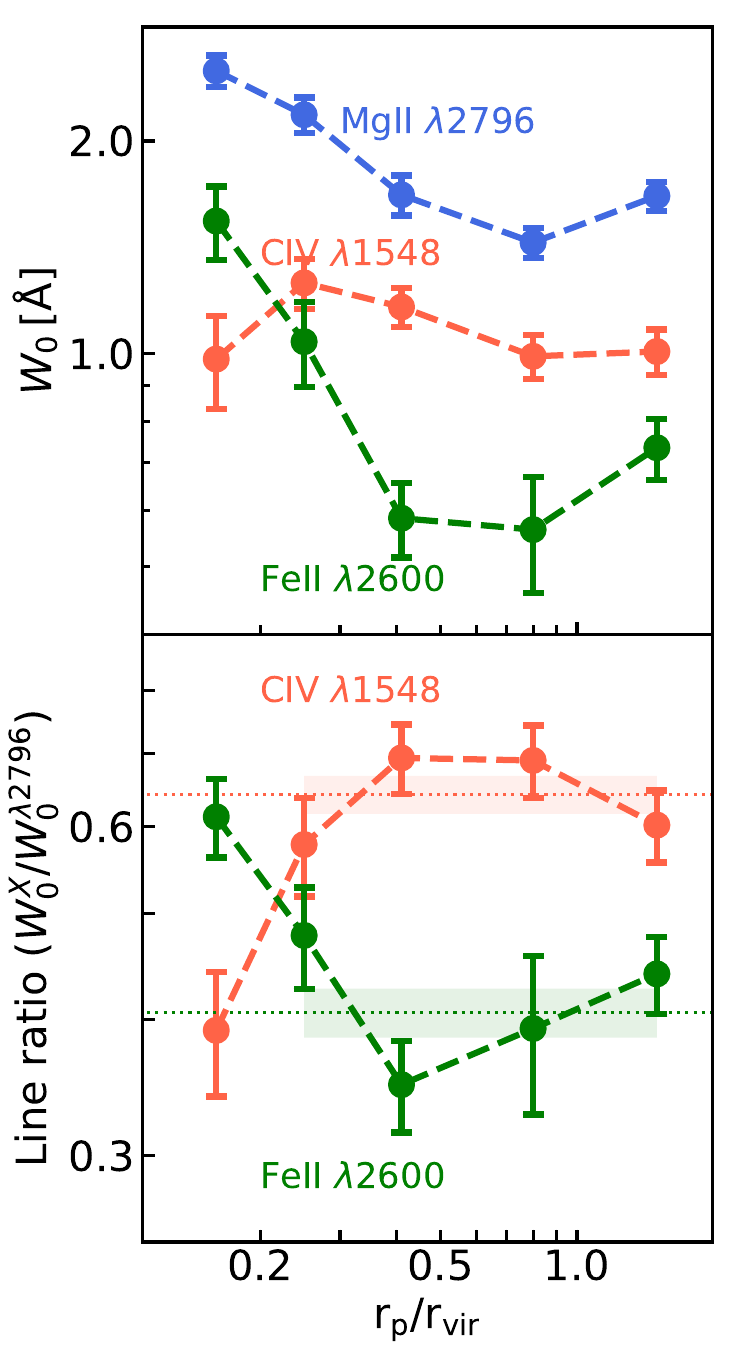}
\caption{Median absorption strengths of MgII$\lambda2796$, FeII$\lambda2600$, CIV$\lambda1548$ and the line ratios as a function of $r_{p}/r_{vir}$ for MgII absorbers with $W_{0}>1 \, \rm \AA$. The upper panel shows the absorption strengths of MgII (blue), FeII (green), and CIV (red). The lower panel shows the absorption strength ratio $\rm FeII/MgII$ (green) and $\rm CIV/MgII$ (red). Uncertainties are estimated by bootstrapping the sample 500 times. In the lower panel, the horizontal lines show the best-fit values for CIV and FeII with $r_{p}/r_{vir}>0.2$ with the uncertainties indicated by the shaded regions.} 
\label{fig:line_ratio_radius}
\end{figure}

\subsection{Absorption Line properties as a function of $r_{p}/r_{vir}$}
\label{text:lineratio}
To connect the relationships of MgII and CIV to galaxies, 
we obtain median composite spectra as a function of $r_{p}/r_{vir}$
at the central velocities of detected MgII absorption lines with $W_{0}^{\lambda2796}> 1 \rm \, \AA$ regardless of the detection of CIV. 
The $W_{0}^{\lambda2796}> 1 \rm \, \AA$ cut is to ensure that we select MgII absorption lines with similar properties across the $r_{p}/r_{vir}$ range. To include as many spectra as possible, we use individual spectra with the quasar spectral S/N around MgII $>3$ and around CIV $>2$. Figure~\ref{fig:composite_spectra} shows the median composite spectra around 6 absorption line species as a function of $r_p/r_{vir}$ indicated by the colors. The shaded colors indicate the uncertainties estimated by bootstrapping the sample 500 time. 

The composite spectra reveal a significant trend. In the most inner region, the absorption line strengths of low-ionized species, including MgI, MgII, FeII, AlII, are stronger than that of the low-ionized species in outer regions, while such a trend is not observed for CIV absorption. 
The absorption line strengths of MgII, FeII, and CIV are shown in the upper panel of Figure~\ref{fig:line_ratio_radius} indicated by blue, green, and red data points respectively. As shown in the figure, MgII absorption strengths increase from outer regions to inner regions by a factor of 1.5. FeII shows a similar trend with a factor of 2.5 increase in the most inner region. In contrast, CIV absorption strength does not change significantly across the impact parameter. 

The lower panel of Figure~\ref{fig:line_ratio_radius} summarizes the relative changes of the absorption line strengths, showing the rest equivalent width line ratios $\rm R_{CIV/MgII}$ $(\rm W_{0}^{\lambda1548}(CIV)/W_{0}^{\lambda2796}(MgII))$ in red and $\rm R_{FeII/MgII}$ ($\rm W_{0}^{\lambda2600}(FeII)/W_{0}^{\lambda2796}(MgII)$) in green. For $\rm FeII/MgII$, $\rm R_{FeII/MgII}$ values are consistent of being $\sim 0.4$ from 0.2 to 2 $\, r_{vir}$, while at $<0.2 \, r_{vir}$, the ratio increases significantly to 0.6. For $\rm CIV/MgII$, $\rm R_{CIV/MgII}$ values are around 0.65 at $>0.2 r_{vir}$ and decrease to 0.4 within $0.2 \, r_{vir}$. The color shaded regions show the best-fit values for $\rm R_{FeII/MgII}$ and $\rm R_{CIV/MgII}$ values $0.2<r_{p}/r_{vir}<2$. The $\rm R_{CIV/MgII}$ and $\rm R_{FeII/MgII}$ values in $<0.2\, r_{vir}$ deviate from the best-fit values by $4.4\sigma$ and $3.8\sigma$ respectively.

We propose that the enhancement of FeII/MgII and the decreasing of CIV/MgII in the inner region are driven by the contribution of cool high density gas that only produce FeII and MgII absorption lines with higher FeII/MgII ratio without contributing any CIV absorption. We discuss this scenario in Section~\ref{sec:CGM_structure}.

\begin{figure}
\center
\includegraphics[width=0.4\textwidth]{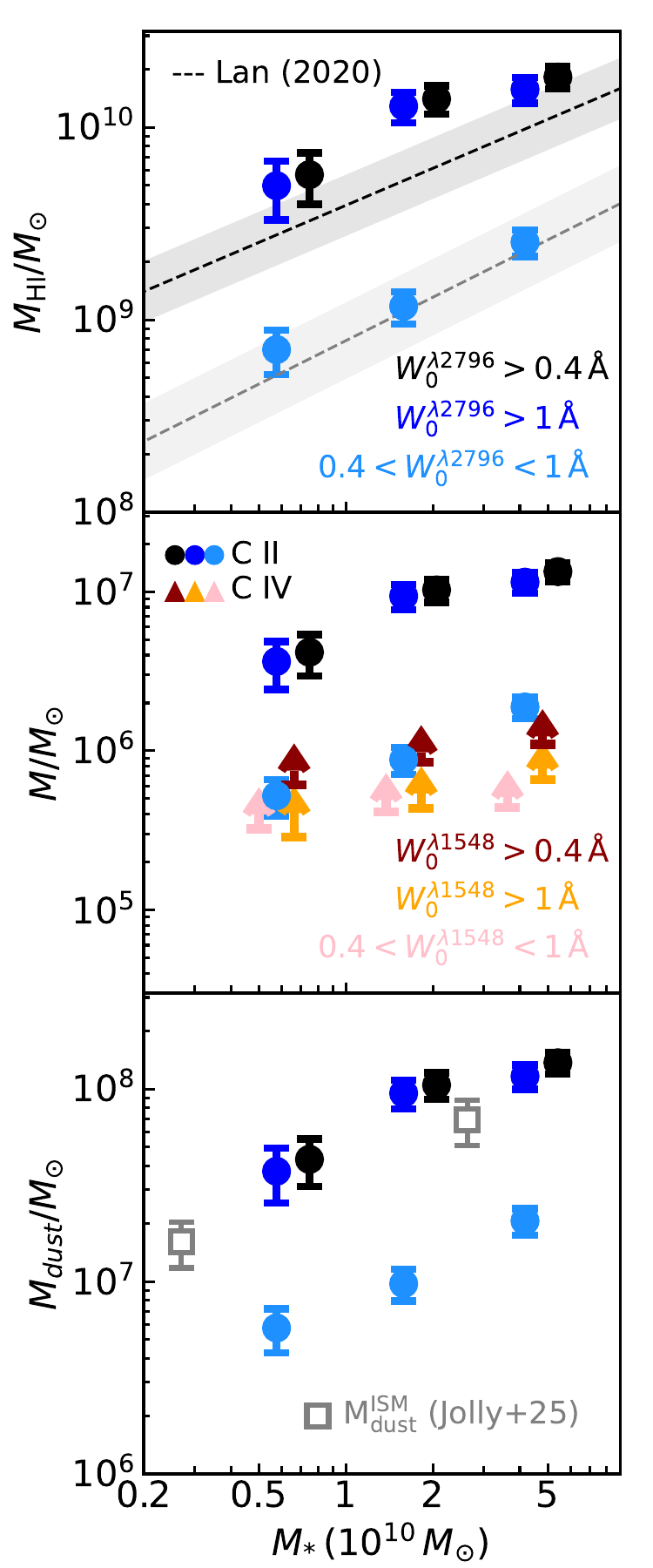}
\caption{Neutral hydrogen, metal, and dust mass in the CGM as a function of stellar mass. The upper panel shows the neutral hydrogen mass traced by weak ($0.4<W_{0}<1 \rm \, \AA$), strong MgII ($W_{0}>1 \rm \, \AA$) absorbers and the sum of the two within $r_{vir}$ with faint blue, blue and black data points respectively.
The dashed lines show the best-fit relations from \citet{lan2020} extrapolated to $z=1.5$. The middle panel shows the estimated Carbon mass traced by weak, strong MgII absorbers and the sum of the two (faint blue, blue and black data points) and weak, strong CIV absorbers, and the sum of the two (pink, orange and red triangles). The measurements from CIV are lower limits given that the CIV lines are saturated. The lower panel shows the dust mass traced by weak, strong MgII absorbers and the sum of the two. The grey square data points show the dust mass in the disk at $z\sim1.4$ from \citet{Jolly25}. Uncertainties are obtained via bootstrapping the samples 500 times.}
\label{fig:mass}
\end{figure}

\section{Discussion}
\subsection{Gas, metal, and dust mass in the CGM}
\textbf{Hydrogen mass ---} Based on the gas distribution, we estimate the mass of neutral hydrogen in the CGM of DESI ELGs traced by MgII. For the neutral hydrogen mass, we adopt the $W_{0}^{\lambda2796}$ and $N_{\rm HI}$ empirical relationship from \citet{lan17},
\begin{equation}
    N_{\rm HI}=10^{18.96}\bigg(\frac{W_{0}^{\lambda2796}}{1 \rm \AA}\bigg)^{1.69}\times(1+z)^{1.88} \rm \, cm^{-2}.
\end{equation}
The relationship is constrained by the data with $W_{0}^{\lambda2796}$ and $\rm N_{HI}$ measurements at $0.1<z<1.6$ \citep{rao06, rao17} and $z\sim3.4$ \citep{matejek13}, reflecting the median amount of $N_{HI}$ traced by MgII absorbers. 
With this relationship, we estimate the total neutral hydrogen mass within the virial radius of DESI ELGs at $z\sim1.5$. The upper panel of Figure~\ref{fig:mass} shows the neutral hydrogen mass traced by MgII absorbers with $W_{0}>1\rm \, \AA$ (blue), $0.4<W_{0}<1\rm \, \AA$ (light blue) and $W_{0}>0.4\rm \, \AA$ (black) as a function of stellar mass. We note that the mass traced by strong absorbers is a factor of 5 - 10 higher than the mass traced by weak absorbers. 
Based on the best-fit redshift evolution from \citet{lan2020} with the measurements at $0.4<z<1.3$, one can predict neutral hydrogen mass at $z=1.5$ traced by MgII absorbers around star-forming galaxies as shown by the black and grey dashed lines. The consistency illustrates that with a fixed stellar mass, the neutral hydrogen mass traced by MgII absorbers around star-forming galaxies continues to increase toward higher redshifts with a trend similar to the redshift evolution of the interstellar medium mass \citep[e.g.,][]{Scoville17}. 
We also note that the neutral hydrogen mass in the CGM is comparable to the neutral hydrogen mass in the ISM ($M_{HI}\sim1-2\times10^{10}\, M_{\odot}$ in $\sim10^{10} M_{\odot}$ star-forming galaxies) measured via 21 cm emission at $z\sim1$ \citep{Chowdhury24}.

\textbf{Carbon mass ---} We also estimate the carbon mass in the CGM. For the carbon traced by MgII absorption lines, we adopt the $W_{0}^{\lambda2796}$ and SiII column density relation from \citet{lan17} based on weak unsaturated SiII line at $1808 \, \AA$ detected in SDSS composite spectra. We then assume a solar relative abundance between C and Si \citep{Asplund21} being $N_{C}/N_{Si}=8.91$ and the same level of depletion (-0.6 dex) \citep[e.g.,][]{lan17}. The estimated carbon masses traced by strong (blue), weak absorbers (light blue) and the sum of the two (black) are shown in the middle panel of Figure~\ref{fig:mass}. Similar to the neutral hydrogen, strong MgII absorbers trace a factor of 5-10 more mass than weak MgII absorbers. 

For the carbon mass traced by CIV absorbers, we estimate the lower limits of the mass given that CIV absorption lines in this work are saturated. We adopt $\rm N_{CIV}\geq 10^{14.3}\, cm^{-2}$ for weak CIV absorbers and $\rm N_{CIV}\geq 10^{14.7}\, cm^{-2}$ for strong CIV absorbers \citep[e.g.,][]{Anand25}. The estimated lower limits are shown in the middle panel of Figure~\ref{fig:mass} with the pink, orange, and dark red data points indicating the measurements for weak, strong CIV absorbers and the sum of the two respectively. 
Based on CLOUDY simulation, \citet{Bordoloi14} estimate the maximum fraction of C in the CIV phase being 0.3. By correcting such a factor, the minimum amount of carbon mass traced by CIV absorbers ($W_{0}>0.4\rm \, \AA$) is $\sim 3\times10^{6} M_{\odot}$.
Together, our results show that the carbon mass is at least $\sim10^{7} M_{\odot}$ for DESI ELGs with stellar masses of $10^{10} \rm \, M_{\odot}$ at $z\sim1.5$. 
To provide some perspective, assuming $\rm 12+log_{10}[O/H]\sim8.5$ metallicity \citep{Zahid14} and solar abundance ratio \citep{Asplund21}, the carbon mass in the ISM of galaxies at $z\sim1.5$ with $M_{*}\sim10^{10} \, \rm M_{\odot}$ is $\sim 6\times10^{7} \, \rm M_{\odot}$. To make a fair comparison, if we further consider the -0.6 dex depletion on the abundance of ISM carbon in the gas phase, the carbon mass is $\sim1.4\times 10^{7} \,\rm  M_{\odot}$, a similar amount as observed in the CGM. This indicates that the CGM contains a non-negligible amount of metals produced by stars at $z\sim1.5$. 

\textbf{Dust mass ---} Finally, we calculate dust mass in the CGM traced by MgII absorbers. We adopt the $W_{0}^{\lambda2796}$ and dust reddening $E_{B-V}$ relation from \citet{menard08} and \citet{menard12}. 
 Following \citet{menard_2010} and \citet{menard12}, we calculate the surface mass density of dust with
\begin{equation}
    \rm \Sigma_{dust} = \frac{ln 10 \, A_{v}}{2.5\, K_{ext,V}}, 
\end{equation}
where the extinction value, $\rm A_{v}$, is based on $E_{B-V}$ with $R_{V}=3.1$, and $\rm K_{ext,V}$ is the extinction-to-dust mass coefficient. We use $\rm K_{ext,V}\simeq 1.54\times10^{4} \, cm^{2} \, g^{-1}$ for the SMC-type dust \citep{menard12, menard08}. The estimated dust masses in the CGM traced by strong and weak MgII absorbers and the sum of the two are shown in the lower panel of Figure~\ref{fig:mass}. We also show the dust mass measured in the ISM of star-forming galaxies at similar stellar mass at $z\sim1.4$ from \citet{Jolly25} (grey data points) \citep[see also][]{Santini14}. The results show that for galaxies with $M_{*}\sim10^{10}\, M_{\odot}$, their CGM and ISM contain similar amount of dust mass, $\sim10^{8} \, M_{\odot}$, indicating that a significant amount of dust produced in galaxies was expelled by galaxies and integrated in the halos. 
We note that the above mass estimations are based on the assumption that the properties of MgII and CIV absorbers around DESI ELGs are consistent with the general MgII and CIV absorber populations detected in random quasar sightlines.

\begin{figure}
\center
\includegraphics[width=0.4\textwidth]{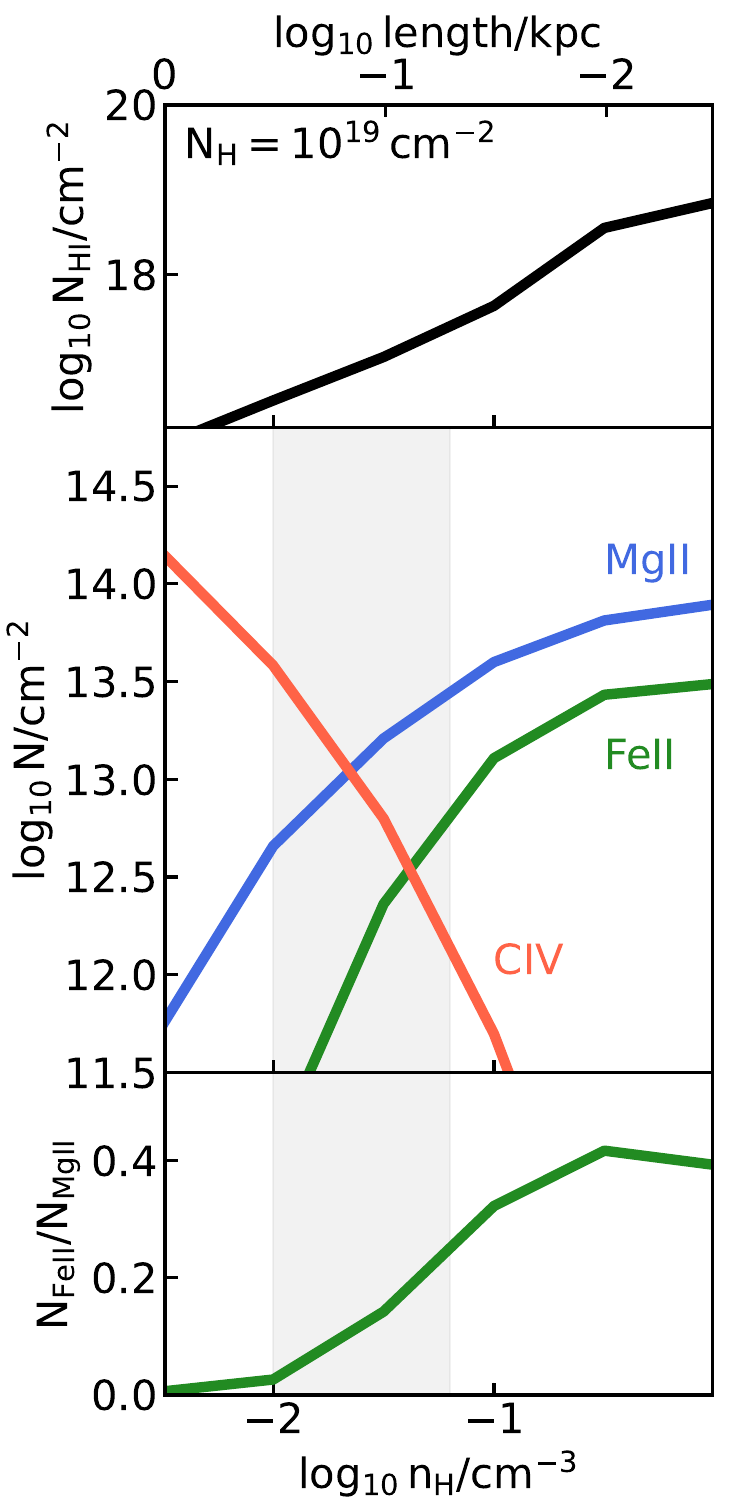
}
\caption{Column density of neutral hydrogen, MgII, FeII, CIV and the ratio between FeII/MgII as a function of gas volume density. \emph{Top:} Neutral hydrogen column density. \emph{Middle:} MgII (blue), FeII (green), CIV (red) column density. \emph{Bottom:} column density ratio of FeII and MgII. } 
\label{fig:cloudy}
\end{figure}

\begin{figure*}
\center
\includegraphics[width=0.95\textwidth]{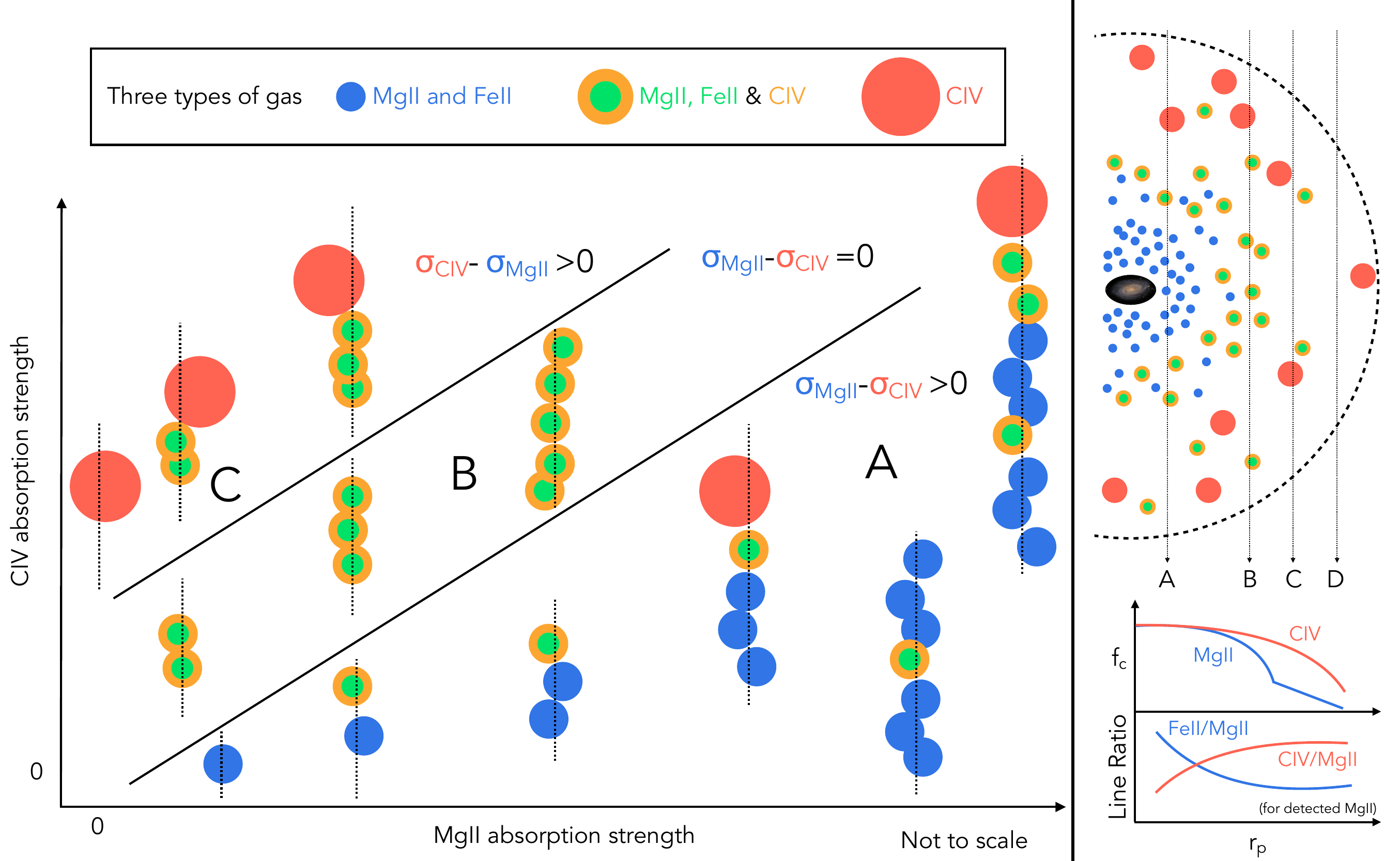}
\caption{Schematic picture for illustrating the contributions of three types of gas observed along the lines of sight. Blue data points represent high density gas producing MgII and FeII absorption lines with high FeII/MgII ratio. The high density gas tends to exist in the inner CGM. Red data points represent low density gas producing CIV absorption lines only. Those gas tends to exist at the outer region of the CGM. Green and orange data points represent gas with intermediate density which can produce MgII, FeII, and CIV with lower FeII/MgII ratio than the blue high density gas. The intermediate density gas tends in exist between the regions producing high density and low density gas. As illustrated in the picture, the observed MgII and CIV absorption strength relation can be explained by intercepting different numbers of these three types of gas. Together with the spatial distribution, one can also explain the covering fraction and the line ratio results shown in Figure~\ref{fig:line_ratio_radius}.} 
\label{fig:picture}
\end{figure*}

\subsection{The structure of the CGM}
\label{sec:CGM_structure}
We now propose a schematic picture of the CGM structure that can explain the observed properties of the CGM traced by MgII and CIV absorption lines. Before doing so, we perform a CLOUDY simulation \citep{cloudy23,ferland98} to illustrate the column density behavior of MgII, FeII and CIV.  The results are shown in Figure~\ref{fig:cloudy}. In this simulation, we consider gas with a fixed total hydrogen column density $10^{19} \, \rm cm^{-2}$, the UV radiation field from \citet{HM05} at $z=1.5$, solar metallicity, and depletion factors for Mg, C, and Fe being -0.6, -0.6, and -1 dex respectively. The metallicity and depletion factors are motivated by column density measurements based on weak unsaturated lines in \citet{lan17}. 
The upper and middle panels of Figure~\ref{fig:cloudy} show the estimated column densities of HI (upper panel), MgII, FeII and CIV (middle panel). The CLOUDY result shows that MgII and FeII column densities decrease with the decrease of the volume density $\rm n_{H}$, while CIV column density shows a reverse trend. Moreover, FeII column density decreases faster than MgII column density toward lower $\rm n_{H}$, resulting in a lower $\rm N_{FeII}/N_{MgII}$ ratio than the ratio at higher $\rm n_{H}$ as shown in the lower panel. The grey shaded region indicates the density region that can produce significant MgII, FeII, and CIV 
absorption lines simultaneously. 

Based on the column density behaviors of MgII, FeII, and CIV, we now introduce three types of gas:
\begin{itemize}
    \item Cool dense gas  ($\rm n_{H}>10^{-1}\, cm^{-3}$): gas clouds in this region have small sizes (10 pc scale), high column densities of HI, MgII and FeII, and $\rm N_{FeII}/N_{MgII}\sim0.4$. 
    \item Cool diffuse gas ($\rm 10^{-2}<n_{H}<10^{-1}\, cm^{-3}$): gas clouds in this region can produce MgII, FeII, and CIV absorption lines with $\rm N_{FeII}/N_{MgII}<0.4$;
    \item Warm diffuse gas  ($\rm n_{H}<10^{-2}\, cm^{-3}$): gas clouds in this region have large sizes (1 kpc scale) can produce high CIV column density with negligible amount of MgII and FeII.  
\end{itemize}
We propose that quasar lines of sight intercepting combinations of these three types of gas yield the observed properties of MgII and CIV. 
Figure~\ref{fig:picture} shows a schematic picture with the blue, green-orange, and red data points representing the above mentioned cool dense gas, cool diffuse gas, and warm diffuse gas respectively. 

First, as shown in Figure~\ref{fig:MgII_CIV_line_width}, with a fixed $W_{0}^{\lambda2796}$ of MgII absorption lines, the CIV absorption line strength varies. One can start with the middle region of the left panel of Figure~\ref{fig:picture} where the sightline intercepts the clouds consisting of the cool diffuse gas. This leads to MgII absorption and CIV absorption with same kinematics structure ($\rm \sigma_{MgII}-\sigma_{CIV}=0$). If we move up along the y-axis, the systems have stronger CIV absorption lines and broader kinematics structure than MgII absorption lines ($\rm \sigma_{CIV}-\sigma_{MgII}>0$). These systems intercept additional gas clouds that produce CIV only. On the other hand, moving down along the y-axis, where $\rm \sigma_{MgII}-\sigma_{CIV}>0$, those systems can only be explained by intercepting gas that primarily produces MgII absorption lines only. 

If we consider absorption lines with increasing MgII absorption strength, one can find that the CIV absorption lines do not increase along with MgII absorption lines as shown in Figure~\ref{fig:MgII_CIV_line_width}. This trend indicates that the increasing MgII absorption strengths and the corresponding kinematics structure are dominated by gas that produces mainly MgII absorption lines visualized in Figure~\ref{fig:picture}, while the sightline can still intercept the cool diffuse gas and the warm diffuse gas given that they have higher covering fractions. Therefore, CIV absorption lines are detected along those sightlines, but do not correlate with MgII absorption lines. 

With this picture, we now further add that the three types of gas preferentially locate at different regions in the CGM as shown in the top right corner of Figure~\ref{fig:picture}.
\begin{itemize}
    \item The cool dense gas that produces MgII and FeII absorption lines mainly concentrates in the inner region, e.g. $r_{\rm 3D}\leq0.2\, r_{vir}$, 
    \item the cool diffuse gas that produces MgII, FeII and CIV locates at the intermediate region, e.g. $0.2<r_{\rm 3D}<0.5\, r_{vir}$,
    \item the warm diffuse gas that primarily 
    produces CIV is mostly at the outer region, e.g. $r_{\rm 3D}>0.5\, r_{vir}$.
\end{itemize}
Sightlines with $r_{p}<0.2 \, r_{vir}$ (A label) intercept these three regions simultaneously with MgII absorption lines preferentially originated from the inner region. This yields detected MgII absorption lines located at the A region shown in the left panel. This can explain the increasing FeII/MgII and decreasing CIV/MgII ratios given that the cool dense gas tends to have high FeII/MgII ratio (as shown in the lower panel of Figure~\ref{fig:cloudy}) and the CIV absorption is not associated with dense components. 
For sightlines at $r_{p}>0.2 \, r_{vir}$ (B label) with detected MgII absorption lines, the MgII, FeII, and CIV are mostly from the cool diffuse gas which yields a lower FeII/MgII ratio and higher CIV/MgII ratio than the ratios in the inner region. 
Beyond $r_{p}>0.5 \, r_{vir}$, the covering fraction of MgII cool gas decreases significantly (below $10\%$), while the covering fraction of CIV gas is $\sim40\%$. In this case, sightlines (C label) tend to intercept CIV gas without any MgII absorption. This again can be explained by the warm diffuse gas preferentially existing at the outer region. Taking together, this CGM structure can explain the observed properties of MgII, FeII and CIV around DESI ELGs at $z\sim1.5$. 

This picture is further supported by studies investigating the general behavior of absorption line systems. For example, using high-resolution spectra, \citet{churchill00} investigated the MgII and CIV connection with 45 sightlines, showed a same pattern of the MgII and CIV relation, and proposed classification scheme based on $W_{0}^{\lambda1548}/W_{0}^{\lambda2796}$. They found that systems in the lower-right corner of Figure~\ref{fig:MgII_CIV_line_width} (high $W_{0}^{\lambda 2796}$ and low $W_{0}^{\lambda1548}/W_{0}^{\lambda2796}$) are mostly associated with damped Lyman-$\alpha$ systems (DLAs) and with higher $W_{0}^{\lambda 2600}/W_{0}^{\lambda2796}$  --- high density gas. \citet{matejek13} also explored the MgII, CIV, and HI relation at $z>2$ and reached to a similar conclusion.
These results indicate that the corresponding mechanisms operate across cosmic time.

We now discuss the possible physical mechanisms and models that can account for the observed properties of the CGM. 

\textbf{Pressure-confined cool gas clouds --- }
Several studies \citep[e.g.,][]{mo96, Maller04, lan19} have considered the possibility that the cool gas in the CGM is in the pressure equilibrium with the surrounding hot gas and is photo-ionized by the cosmic UV background radiation. While the hot gas density and temperature profiles around galaxies remain to be constrained precisely, the current observational constraints \citep[e.g.,][]{miller13} and model predictions \citep[e.g.,][]{Maller04, Barbani23} indicate that the pressure is $\rm \sim10^{3-4} \, K/cm^{3}$ in the inner halo and $\rm \sim10^{2}\, K/cm^{3}$ on the outskirt for $\rm 10^{12} \, M_{\odot}$ halos. This range of pressure yields the corresponding cool gas ($10^{4}$ K) cloud volume densities being $\rm 0.1-1 \, cm^{-3}$ in the inner region to $\rm 0.01 \, cm^{-3}$ at the outskirt. In other words, this pressure confinement scenario can possibly produce our proposed three types of gas clouds and observed properties in the CGM shown in Figure~\ref{fig:picture}, including the line ratio, the preferred 3D location, and characteristic sizes. This supports the picture that MgII, FeII, and CIV absorption lines observed in the halos are given rise by pressure confined and photo-ionized gas clouds. We emphasize that Figure~\ref{fig:cloudy} only shows the column densities of a single cloud, while detected absorption line systems typically consist of many clouds ($\rm N_{cloud}\gtrsim10$ for $W^{\lambda2796}_{0}>1 \,\rm  \AA $ systems) \citep[e.g.,][]{Churchill03, churchill20}. The combination of multiple clouds can yield the HI column density and metal column density \citep[e.g.,][]{hummels24, bisht25, yang25} being similar to the observed column densities in MgII and CIV absorbers \citep[e.g.,][]{lan17, Anand25}.

\textbf{Hierarchical density structure of the CGM --- }
\citet{stern16} proposed that the circumgalactic gas has a hierarchical density structure with cool and high-density gas clouds embedded in large warm and low-density clouds. With this assumption, their model can fit the observed column densities of multiple absorption line species tracing different density and temperature of the CGM simultaneously at $z\sim0$. The major difference between our proposed picture and the model in \citet{stern16} is that in our picture, the cool dense gas in the inner halo exists without any associated warmer gas, while cool dense gas clouds are assumed to be always surrounded by diffuse gas in \citet{stern16}. Based on their assumption, when the absorption line systems tracing cool gas, such as MgII, are detected, associated absorption lines tracing warmer gas, such as CIV, are expected to be detected as well. However, observationally, systems with strong MgII absorption lines and with weak or no CIV absorption lines with kinematics misalignment between the two are detected (systems at the lower right region in Figure~\ref{fig:MgII_CIV_line_width}), suggesting that cool dense gas and warm diffuse gas do not always coexist. Nevertheless, we note that in \citet{stern16}'s model, multiple high-density gas clouds can exist in one low-density gas cloud. In this case, it is possible that a sightline intercepts many high-density gas clouds in one low-density gas cloud. This is expected to give rise to relative strong low-ionized absorption line with respect to the associated high-ionized absorption line which can partially explain the variation of observed MgII and CIV relation. We also note that the kinematics alignment of different absorption line species, another indicator for the association of absorption lines, is not considered in the model of \citet{stern16}.

\textbf{Turbulence-dominated CGM ---} 
Based on the FIRE simulations, \citet{Stern21} proposed that the inner CGM region ($\rm <0.3\, r_{vir}$), where cooling time is shorter than the free-fall time, maintains cool without being virialized to the hot temperature. Under this scenario, the inner CGM is dominated by turbulent, volume-filling cool gas rather than pressure-confined cool "clouds". For $\rm 10^{12} \, M_{\odot}$ halos, this inner cool CGM region disappears toward lower redshifts, producing a redshift evolution of the cool gas around galaxies similar to the redshift evolution of strong MgII absorbers globally \citep[e.g.,][]{zhu13} as well as the redshift trend shown in Figure~\ref{fig:MgII_CIV_rvir}. \citet{kakoly25} further produce mock absorption lines induced by this cool inner CGM and estimate the rest equivalent widths of multiple absorption lines around $10^{12} \rm \, M_{\odot}$ halos. 
Their results show that the average MgII rest equivalent widths evolve with redshifts similar to the observed trend, indicating that this scenario offers a promising mechanism to explain the redshift evolution of strong MgII absorbers. However, we note that in comparison with our CGM measurements at $z\sim1.5$ at $r_{p}<0.2 r_{vir}$, the estimated MgII rest equivalent width ($W_{0}^{\lambda2796}\sim0.8 \rm \AA$) is significantly lower than the observed strength ($W^{\lambda2796}_{0}\sim 2 \rm \AA$) by a factor of 2-3.  This difference can be due to that the amount of cool gas with required density to produce high MgII column density is not sufficient in the simulation.

\section{Conclusions}
Utilizing the latest DESI dataset, we construct a large sample of ELG-quasar pairs at $z\sim1.5$ and measure the properties of the cool and warm gas around galaxies traced by MgII and CIV absorption lines. We investigate the relationship between the properties of multiphase gas and that of DESI ELGs and how the properties of gas traced each other. Our main findings are summarized as follows:
\begin{enumerate}
    \item We find that MgII and CIV have distinct average absorption line profiles around DESI ELGs. The MgII absorption strengths decrease with impact parameters faster than the CIV absorption strengths. This indicates that cool MgII gas and warm CIV gas have different distributions around galaxies. 
    \item We find that both the Mg II and C IV covering fractions in physical space correlate with the stellar mass of DESI ELGs, with significances of $\sim4\sigma$ and $\sim2\sigma$ (tentative) respectively, while there is no significant correlation with SFR over the SFR range probed by our sample.
    \item When considering the gas distributions with respect to the sizes of the dark matter halos, the gas distributions around ELGs with different mass align consistently with each other. This behavior is seen for both the cool and warm gas traced by MgII and CIV absorption lines, indicating that the gas distributions primarily link to the gravitational potential of the systems.
    \item The gas velocity dispersions of both MgII and CIV are $\sim 100$ $\rm km \, s^{-1}$ within the halos, consistent with the velocity dispersions of dark matter particles, while the values reach to $\sim 200$ $\rm km \, s^{-1}$ beyond the halos possibly contributed by the gas associated with nearby galaxies (two-halo term). 
    \item We explore the connections between MgII and CIV gas and show the two are not tightly coupled. With a fixed property of one species, such as $W_{0}$ and $\sigma_{gas}$, the property of the other species varies by several-fold. 
    \item We measure the median $W_{0}$ of MgII, FeII and CIV of detected strong MgII absorbers around ELGs and find that at $<0.2\times r_{vir}$, 
    the FeII/MgII ratio is elevated and the CIV/MgII ratio is suppressed compared with the measurements at larger scales. Both signals are detected with $\sim4\sigma$. 
    
    \item We propose a schematic picture of the CGM structure, illustrating that multiphase gas, which is not co-spatial but detected along the sightlines, can explain the observed line ratio behaviors and the MgII–CIV connections.
    
    \item Finally, combining measurements from literature, we 
    show that the covering fraction and rest equivalent width profiles of MgII and CIV evolve with redshifts. We also estimate the masses of neutral hydrogen, metals, and dust in the CGM, finding amounts comparable to those in the ISM.

\end{enumerate}
These results provide novel and precise measurements on the properties of multiphase CGM of star-forming galaxies at $z\sim1.5$, offering key tests on the mechanisms that regulate galaxy evolution and shape the CGM structure.

The results of this work also demonstrate that with ongoing and upcoming large cosmological surveys, including DESI-II and Spec-5 \citep{desiiispec5}, PFS \citep{pfs14}, Euclid \citep{euclid22}, and Roman Spectroscopic Survey \citep{roman22}, the multiphase CGM and its connection to galaxy properties at $z>2$ can be measured with unprecedented precision. High-resolution imaging from space telescopes, such as Euclid \citep{euclid22} and Roman Telescope, will further allow the characterization of gas distributions linking to the morphology of galaxies. Combined with existing measurements at lower redshifts, these datasets will map galaxy–multiphase CGM connections across cosmic time, from $z\sim3-4$ to the present.
Together, the joint constraints on galaxy and CGM properties will shed new light on the physical mechanisms of galaxy evolution and advance our understanding of the cosmic baryon cycle.

\section*{Data Availability}
All data points shown in the figures are available at \href{https://doi.org/10.5281/zenodo.17284206}{Zenodo}.
This deposit also includes a Jupyter notebook that includes the code
linking the provided data files to the published figures.

\begin{acknowledgments}
We dedicate this work to the memory of \newline Jacqueline Bergeron, whose pioneering contributions laid the foundation for this field. Her vision and legacy continue to guide and inspire our research. We thank Lucas Napolitano Simon Weng, Jonathan Stern and Aharon Kakoly for their comments and suggestions.
TWL acknowledges supports from National Science and Technology Council (MOST 111-2112-M-002-015-MY3, NSTC 113-2112-M-002-028-MY3), Yushan Fellow Program by the Ministry of Education (MOE) (NTU-110VV007, NTU-111V1007-2, NTU-112V1007-3, NTU-113V1007-4, and NTU-114V1007-5), and MOE-115-YSFMS-0003-001-P2 (NTU 115V2020-1).

This material is based upon work supported by the U.S. Department of Energy (DOE), Office of Science, Office of High-Energy Physics, under Contract No. DE–AC02–05CH11231, and by the National Energy Research Scientific Computing Center, a DOE Office of Science User Facility under the same contract. Additional support for DESI was provided by the U.S. National Science Foundation (NSF), Division of Astronomical Sciences under Contract No. AST-0950945 to the NSF’s National Optical-Infrared Astronomy Research Laboratory; the Science and Technology Facilities Council of the United Kingdom; the Gordon and Betty Moore Foundation; the Heising-Simons Foundation; the French Alternative Energies and Atomic Energy Commission (CEA); the National Council of Humanities, Science and Technology of Mexico (CONAHCYT); the Ministry of Science, Innovation and Universities of Spain (MICIU/AEI/10.13039/501100011033), and by the DESI Member Institutions: https://www.desi.lbl.gov/collaborating-institutions.

The DESI Legacy Imaging Surveys consist of three individual and complementary projects: the Dark Energy Camera Legacy Survey (DECaLS), the Beijing-Arizona Sky Survey (BASS), and the Mayall z-band Legacy Survey (MzLS). DECaLS, BASS and MzLS together include data obtained, respectively, at the Blanco telescope, Cerro Tololo Inter-American Observatory, NSF’s NOIRLab; the Bok telescope, Steward Observatory, University of Arizona; and the Mayall telescope, Kitt Peak National Observatory, NOIRLab. NOIRLab is operated by the Association of Universities for Research in Astronomy (AURA) under a cooperative agreement with the National Science Foundation. Pipeline processing and analyses of the data were supported by NOIRLab and the Lawrence Berkeley National Laboratory. Legacy Surveys also uses data products from the Near-Earth Object Wide-field Infrared Survey Explorer (NEOWISE), a project of the Jet Propulsion Laboratory/California Institute of Technology, funded by the National Aeronautics and Space Administration. Legacy Surveys was supported by: the Director, Office of Science, Office of High Energy Physics of the U.S. Department of Energy; the National Energy Research Scientific Computing Center, a DOE Office of Science User Facility; the U.S. National Science Foundation, Division of Astronomical Sciences; the National Astronomical Observatories of China, the Chinese Academy of Sciences and the Chinese National Natural Science Foundation. LBNL is managed by the Regents of the University of California under contract to the U.S. Department of Energy. The complete acknowledgments can be found at https://www.legacysurvey.org/.

Any opinions, findings, and conclusions or recommendations expressed in this material are those of the author(s) and do not necessarily reflect the views of the U. S. National Science Foundation, the U. S. Department of Energy, or any of the listed funding agencies.

The authors are honored to be permitted to conduct scientific research on I'oligam Du'ag (Kitt Peak), a mountain with particular significance to the Tohono O’odham Nation.
\end{acknowledgments}

\software{Astropy \citep{astropy13,astropy18,astropy22}, Numpy \citep{numpy}, Scipy \citep{scipy}, Matplotlib \citep{Hunter2007}, KapteynPackage \citep{kmpfit}}

\appendix

\section{Completeness Correction}
\label{sec:com}
\begin{figure*}
\center
\includegraphics[width=0.95\textwidth]{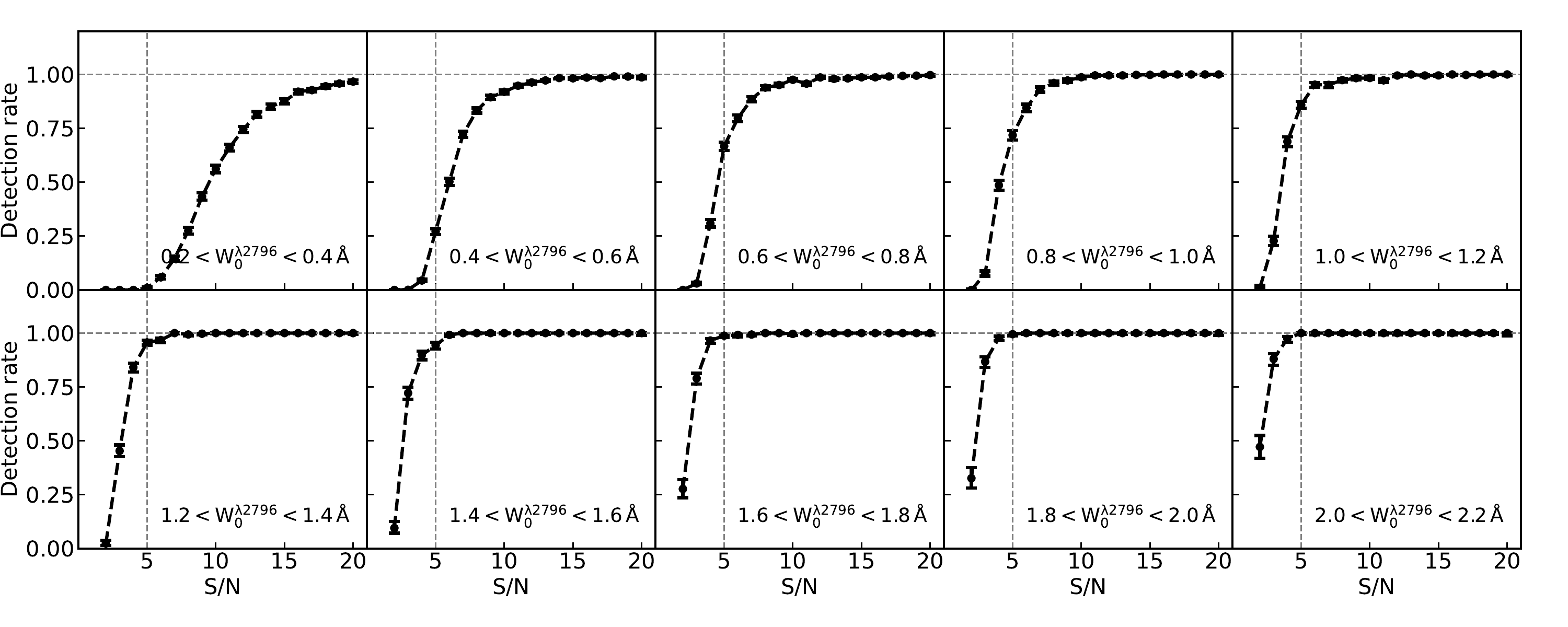}
\includegraphics[width=0.95\textwidth]{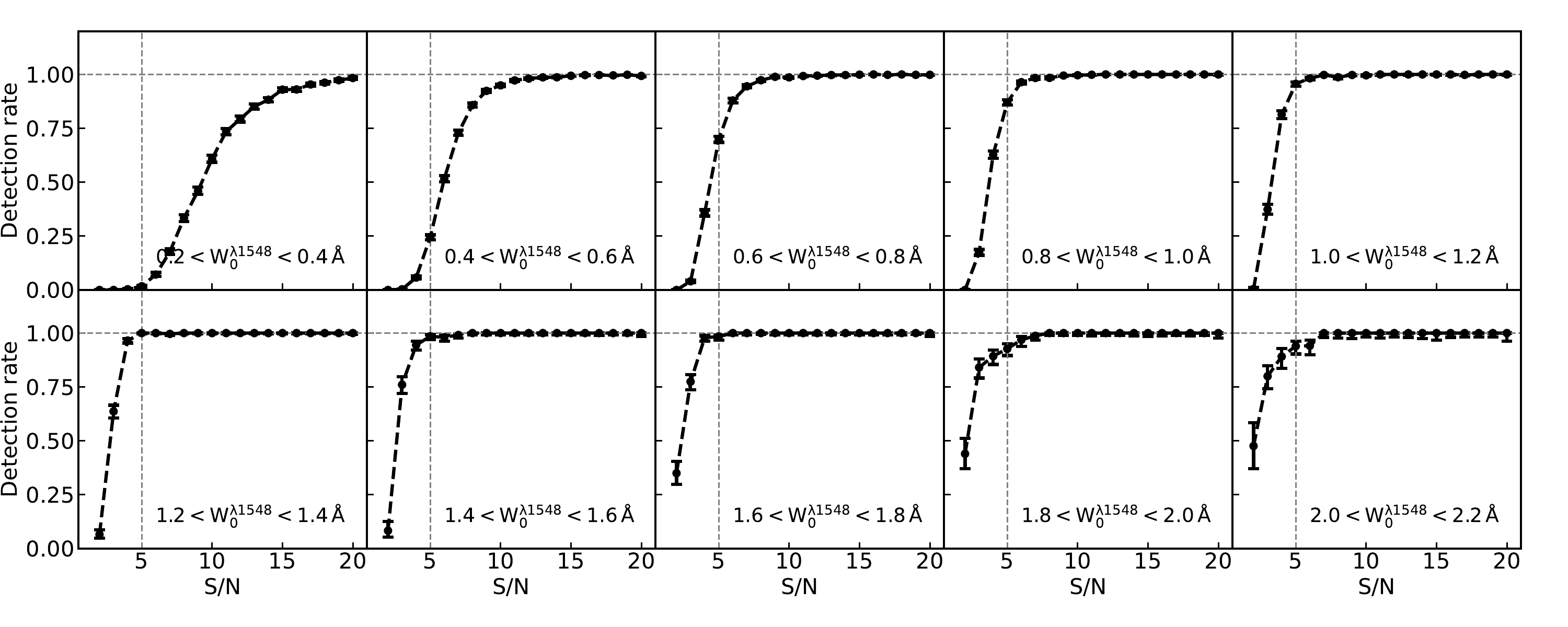}
\caption{Detection rates of MgII (upper panels) and CIV (lower panels) as a function of $W_{0}$ and S/N of the continuum based on mock spectra. The $S/N=5$ is shown by the vertical dashed line.} 
\label{fig:completeness}
\end{figure*}

We perform mock simulations to estimate the detection rate of absorption line systems as a function of the absorption line strengths and the signal-to-noise ratios (S/N) of the continuum. To this end, we randomly select the absorption line systems that are detected in the DESI quasar spectra, use their best-fit absorption properties and create the corresponding absorption line profiles. We then insert the profiles into normalized spectra with a uniform distribution of continuum S/N from 2 to 20. In total, we create 100,000 mock absorption spectra for both MgII and CIV and run the MgII and CIV line detection algorithm that we apply to the real quasar spectra and calculate the detection rate of MgII and CIV absorption lines as a function of their strengths and S/N. Figure~\ref{fig:completeness} shows the results. As expected, the detection rate increases toward high rest equivalent width and high S/N. For our default S/N cut (S/N$>5$), the detection rates of strong MgII and CIV absorbers with $W_{0}>1 \rm \,\AA$ are almost $100\%$ except the systems at the boundary ($W_{0}\simeq1 \rm \,\AA$ and $S/N=5$). For weaker absorption lines, the detection rate decreases. For $0.4<W_{0}<0.6 \rm \,\AA$ and $S/N=5$, the detection rate is $\sim25\%$. To account for the undetected systems due to the S/N of the dataset, we assign a weight, $w_{i}$, for a detected absorber as
\begin{equation}
    w_{i}=\frac{1}{\rm Detection \, rate(W_{0},S/N)}.
\end{equation}
This weight is used for the covering fraction calculation in Eq. 3.

\section{Composite spectra}
\label{apen:composite}
\begin{figure*}
\center
\includegraphics[width=1\textwidth]{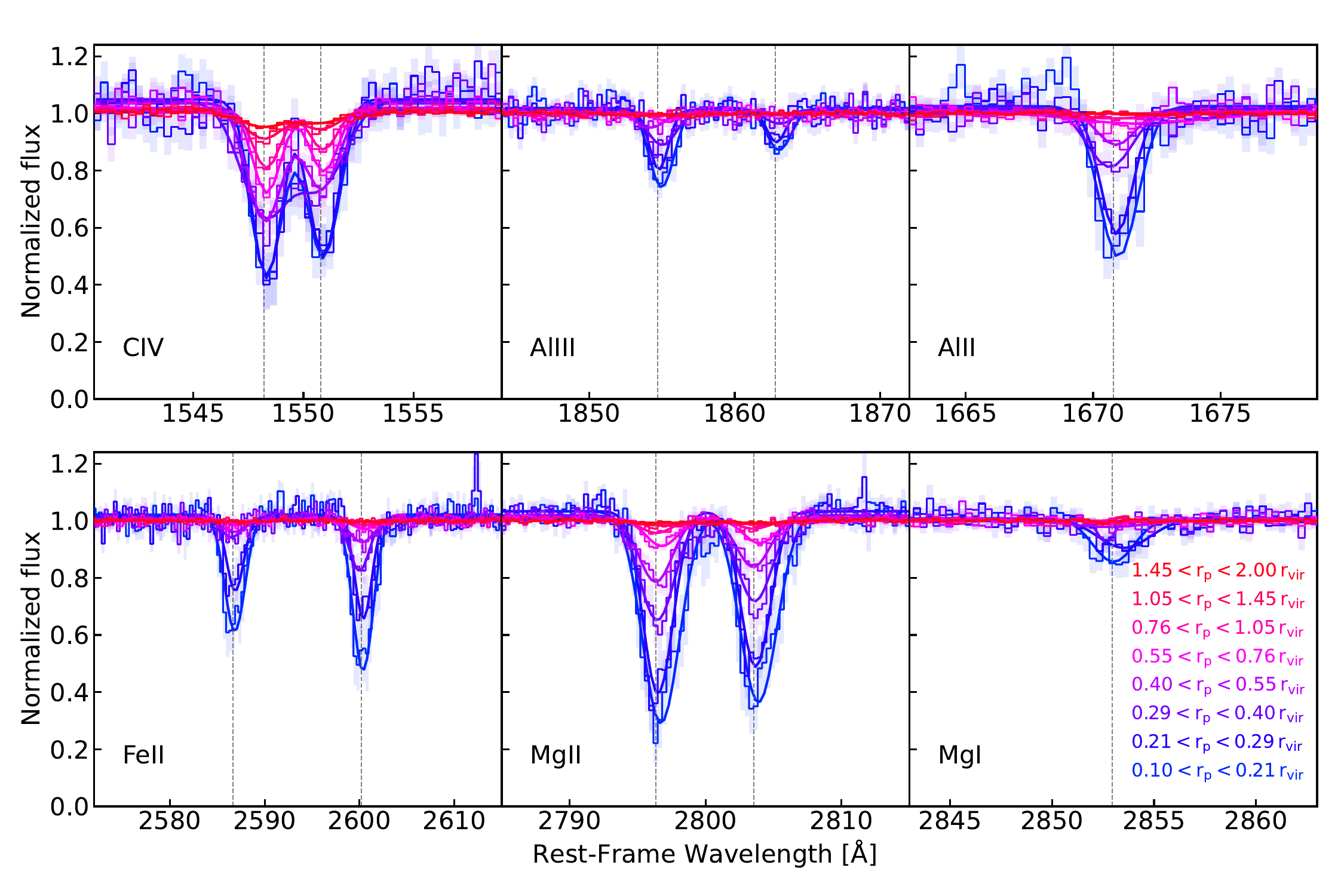}
\caption{
Mean composite spectra around ELGs as a function of $r_p/r_{vir}$. 
Each panel shows the spectral region of an absorption species listed in the lower left corner. The colors indicate the selected $r_p/r_{vir}$ bins and the shaded regions show the uncertainty based on bootstrapping the sample 500 times. 
} 
\label{fig:all_composite}
\end{figure*}

\begin{figure*}
\center
\includegraphics[width=1\textwidth]{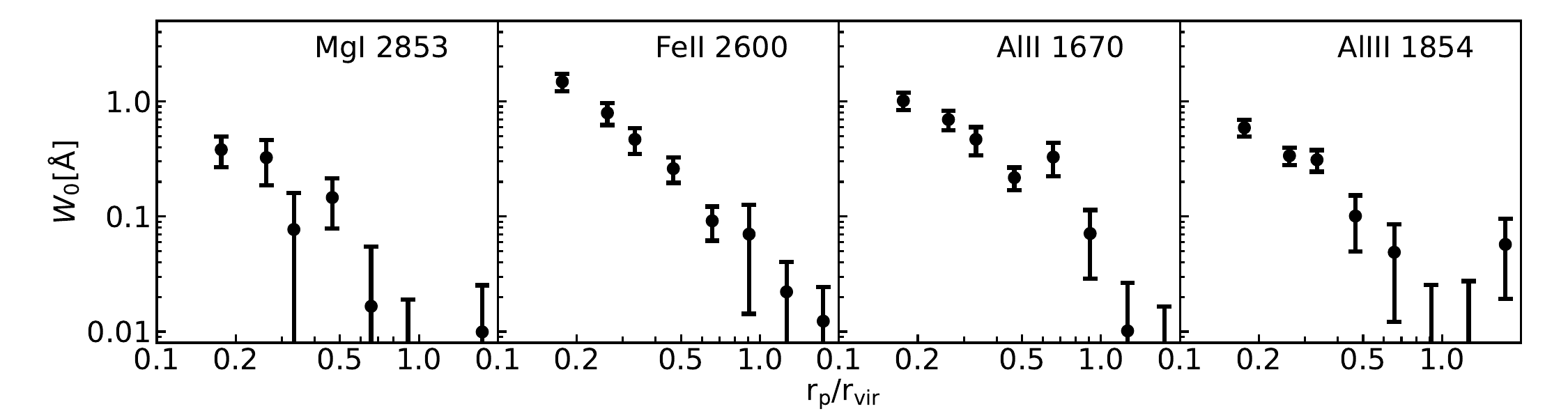}
\caption{
Mean rest equivalent widths of MgI, FeII, AlII, and AlIII (from left to right) as a function of  $r_p/r_{vir}$. The uncertainties are based on bootstrapping the sample 500 times. 
} 
\label{fig:all_mean}
\end{figure*}

In addition to detecting individual MgII and CIV absorption lines in the background quasar spectra as shown in Figure 1, we perform stacking analysis to obtain mean composite spectra for all sightlines with $\rm S/N>3$ as a function $r_{p}/r_{vir}$. The results are shown in Figure~\ref{fig:all_composite} with colors indicating the $r_{p}/r_{vir}$ ranges and the solid smooth lines are the best-fit Gaussian profiles. Figure~\ref{fig:all_mean} shows the best-fit rest equivalent widths of MgI, FeII, AlII, and AlIII, offering absorption line information other than MgII and CIV shown in Figure 1.

\newpage

\bibliography{bibtex.bib}{}
\end{document}